\documentclass[9pt,prx,a4paper,final,english,twocolumn,nobibnotes,aps3,longbibliography]{revtex4-2}
\usepackage{dcolumn}
\usepackage{microtype}
\usepackage{graphicx}
\usepackage{xcolor}
\usepackage{booktabs}
\usepackage[normalem]{ulem}
\usepackage{enumitem}
\definecolor{darkblue}{rgb}{0,0,.65}
\definecolor{bg}{rgb}{0.9,0.95,1}

\usepackage[
    pdfauthor   = {Ophelia Evelyn Sommer},
    pdfstartview= FitH,
    bookmarks   = true,
    colorlinks  = true,
    anchorcolor = black,
    citecolor   = blue,
    filecolor   = red,
    menucolor   = black,
    urlcolor    = blue,
    linkcolor   = blue,
]{hyperref}
\usepackage{natbib}

\usepackage{amsmath,amssymb,amsthm}
\usepackage{bm}
\usepackage{siunitx}
\usepackage{braket} %For bra, ket notation
\newcommand{\Z}{\mathbb{Z}}
\newcommand{\R}{\mathbb{R}}

\newcommand{\hexagon}{%
  \tikz[scale=0.1,baseline=-0.25ex]{\draw (0,0) \foreach \x in {0,60,...,300} { -- ++(\x:1) } -- cycle;}%
}

\DeclareMathOperator{\tr}{Tr}

\let\Im\relax
\DeclareMathOperator{\Im}{Im}

\newcommand{\e}{\mathrm{e}}
\renewcommand{\i}{\mathrm{i}}
\newcommand{\dd}{\mathrm{d}}
\newcommand{\lB}{\ell_B}

\newcommand{\vb}[1]{\mathbf{#1}}
\newcommand{\mr}[1]{\mathrm{#1}}
\newcommand{\mc}[1]{\mathcal{#1}}

\theoremstyle{definition}

\usepackage{tikz-feynman}
\usepackage{tikz-3dplot}
\usepackage{tikz}
\usetikzlibrary{calc}
\usetikzlibrary{shapes.callouts}
\usetikzlibrary{decorations.markings}
\usetikzlibrary{3d}
\tikzset{
    level/.style = {
        ultra thick,
        black,
    },
    connect/.style = {
        dashed,
        gray
    },
    notice/.style = {
        draw,
        rectangle callout,
        callout relative pointer={#1}
    },
    label/.style = {
        text width=2cm
    },
      trans/.style={thick,<->,shorten >=2pt,shorten <=2pt,>=stealth}
}

\begin{document}
\title{Ideal Optical Flux Lattices}

\author{Ophelia Evelyn Sommer}
\affiliation{Department of Physics, Harvard University, Cambridge, MA 02138, USA}
\author{Nigel R. Cooper}
\affiliation{T.C.M. Group, Cavendish Laboratory, University of Cambridge, J.J. Thomson Avenue, Cambridge CB3 0US, United Kingdom \looseness=-1}

\begin{abstract}
The realization of fractional quantum Hall (FQH) states in cold atomic gases is a long-standing goal in quantum simulation. Established approaches, including rapidly rotating gases and tight-binding lattices, are often hampered by low interaction energies and small many-body energy gaps. While optical flux lattices (OFLs) can achieve higher effective magnetic flux densities, standard two-state configurations generate highly non-uniform fields, and extensions to multi-state systems introduce significant experimental complexity. Here, we present a new paradigm for engineering robust FQH phases in OFLs using only two internal atomic states. We show that the introduction of an additional scalar potential provides a generic mechanism for creating Chern bands that are simultaneously essentially flat and ``ideal.'' These desirable properties arise by tuning lattice parameters to certain $N$-flat manifolds $(N=1,2,\dots)$, where the $1$-flat manifold shares its origin with certain ``magic-angle'' conditions familiar from moiré materials.
{A central result is the design of a dark-state OFL whose adiabatic Hamiltonian is \emph{exactly} of Aharonov-Casher (AC) form. This exact AC equivalence guarantees perfectly flat, exactly vortexable Chern bands in the adiabatic limit.} This method allows for precise tuning of band flatness and stabilizes both Abelian and non-Abelian FQH phases. Our scheme is compatible with existing experimental capabilities using vector polarizability, opening practical routes to exploring strongly correlated topological physics with cold atoms.
\end{abstract}

\maketitle

\section{Introduction}

One of the long-standing challenges in quantum simulation using cold
atomic gases is the realization of fractional quantum Hall (FQH) states.
The possibility of using experiments on cold atomic gases to explore FQH states of
bosons -- in place of the fermions familiar from electronic matter --
was first discussed in the context of Bose-Einstein condensates in
harmonic trapping potentials, for which Landau level wavefunctions
emerge under rapid rotation\cite{Cooperadvances,fetterreview}.
Remarkably, the bosonic Laughlin state can be the {\it exact}
ground state at sufficiently high angular momentum\cite{WilkinGS98},
due to the short-range nature of typical interatomic interactions.
Subsequent theoretical work established the phases of short-range
interacting bosons as a function of the Landau level filling
factor, $\nu$\cite{CooperWG00}. In addition to conventional
hierarchy/composite fermion states, bosons form unusually stable
non-Abelian phases: including the Moore-Read phase\cite{MooreR91}
at $\nu=1$\cite{CooperWG00,RegnaultJ03}, and the Read-Rezayi
phases\cite{ReadR99,RezayiRC05}. Achieving these FQH phases in
experiments using rotating gases is extremely challenging, in part
due to the low flux densities and hence low interaction energies.
However, this may be mitigated by tight confinement using optical
tweezers\cite{LuntHRPGJ24}.

One way to increase the interaction energies of FQH states is
to place the atoms in tight-binding lattices, engineered (by
breaking time-reversal symmetry) to host narrow Chern bands\cite{Cooperadvances}. The
Harper-Hofstadter model supports the $\nu=1/2$ bosonic Laughlin
state\cite{sorensen2005fractional} and $\nu=1$ Moore-Read state\cite{palm2021}, as well as other FQH states
that are stabilized by the lattice itself and do not have continuum
analogues\cite{MollerC09}. Interaction energies are boosted compared to
those in rotating gases, but they typically remain small. The state of
the art experiments\cite{leonard2023realization} have shown Laughlin
correlations for two particles, with a many-body energy gap of order
$\sim 1\,\si{\hertz}$.

A separate approach to forming Chern bands for cold atoms uses
the possibility to couple internal (electronic or magnetic)
levels of atoms, to form so-called ``optical flux lattices"
(OFLs)\cite{cooper2011optical}. The main conceit of OFLs is that they
require state-dependent potentials that do not commute at different
points in space. In the deep-lattice limit, the energetically preferred
internal state at each point in real space forms a texture with a
varying direction. The Berry phase associated to this texture becomes an
Aharonov-Bohm phase for an emergent magnetic field associated with the
residual dynamics, that often persists even in the intermediate-shallow
lattice regime.  Thus, OFLs stand in stark contrast to deep-lattice
tight-binding models, in that they are generally shallow lattice
continuum models. As a result, they can operate at higher flux densities
boosting the densities at which FQH states can form, and thereby also
the associated energy scales.

While the emergent magnetic field in OFLs has a nonzero average, it
is typically spatially varying, so these lattices cannot directly
reproduce the uniform continuum Landau levels. In particular, if only two
internal levels are coupled the local magnetic field must vanish
within the unit cell\cite{cooper2011optical}, so the field must
have very large spatial variations. To mimic continuum Landau
levels more closely, OFLs have been proposed using $N>2$ internal
states\cite{cooper2012designing}, which converge quickly to the
continuum Landau levels with increasing $N$. However, experimental
implementations\cite{cooper2013reaching,LauriaKCB24} quickly become
more cumbersome, requiring multiple frequencies of the light field to
control the various interspecies couplings, which can lead to enhanced
losses through light scattering.

In this paper, we take a different approach, and show how OFLs using just two
internal states can be designed to host robust Abelian and non-Abelian
FQH phases.  We draw on theoretical work inspired by advances in two-dimensional electronic platforms, such as the moiré materials
formed from graphene or TMDs. The prevalence of non-trivial band geometry in these platforms has led to a deeper understanding
of the general properties of a single-particle band that are desirable in order to stabilize FQH states\cite{wang2021exact,ledwith2023vortexability,estienne2023ideal,TarnopolskyPRL2019MATBG,liu2025theory,FujimotoPRL2025HighVortex,LedwithPRR2020TBG,ChongPRL2024FCITMD,DongPRL2023CFL,ReddyPRB2023FCITMD,JacksonNatCom2015GeoStab,ClasssenPRL2015PMDuality,BrunoPRB2021FlatChern,TomokiPRB2021TraceCondnLLL,MoralesPRL2024MATTMD}. A prominent notion emerging from these considerations is that of
``ideal'' bands, which are ``vortexable'', and consequently allow for
the appearance of exact FQH states for short-range interactions,
generalizing the Laughlin state in the lowest Landau level to exact
states in these bands.

Our principal result is to present a
generic mechanism for creating essentially flat and ideal Chern $|C|=1$
bands, upon introducing an additional scalar potential to OFLs. This
additional scalar potential can be used to tune the flatness of the
band to a high precision. These flat bands host fractional quantum Hall
phases in the presence of repulsive interactions, and can be realized
with currently existing experimental tools in optical lattices using
vector polarizability.

{A central result within this framework is the construction of a dark-state OFL whose adiabatic Hamiltonian is \emph{exactly} of Aharonov-Casher (AC) form (see Sec.~\ref{sec:darkofl}). While the AC Hamiltonian has served as a theoretical ideal for flat bands~\cite{aharonovGroundStateSpin1/21979,MoralesPRL2024MATTMD,shiAdiabaticApproximationAharonovCasher2024}, prior realizations have only approximated this structure. The dark-state OFL constructed here achieves it exactly in the adiabatic limit, yielding perfectly flat, exactly vortexable Chern bands as an analytical result rather than a numerical observation. This constitutes a physically realizable benchmark for ideal band physics, and offers a tunable, well-characterized platform in which the AC limit is achieved precisely.}

The outline of the paper is as follows. In Sec.~\ref{sec:ofl} we provide background on optical flux lattices, emphasizing their description as reciprocal space tight-binding models. We then describe the concepts of ideality in Sec.~\ref{sec:ideal}, and explain how to use these to design ideal OFLs. We provide explicit examples of these design principles in Sec.~\ref{sec:examples}, including also for dark-state optical flux lattices. In Sec.~\ref{sec:strongly}, we demonstrate their utility for stabilizing Abelian and non-Abelian FQH states.

\begin{figure}
	\centering
	\includegraphics{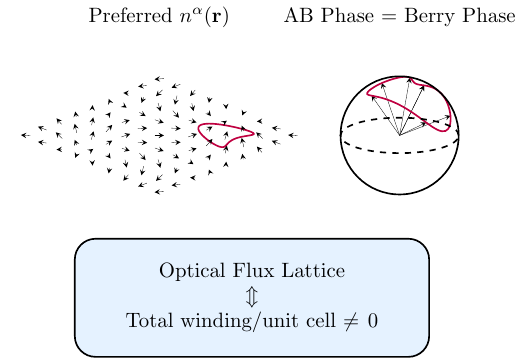}
	\caption{Graphical illustration of the connection between the low-energy spinor texture and the effective magnetic field. The Aharonov-Bohm (AB) phase picked up upon parallel transport of the emergent scalar particle is the Berry phase of the spinor texture.} \label{fig:adiabatic_spintexture}
\end{figure}

\section{Background on Optical Flux Lattices}

\label{sec:ofl}

\subsection{Real Space Viewpoint}

Optical lattices are created by
coupling atoms to lasers that transfer
momentum. For optical flux lattices, we further allow the lasers to change the internal state of the atom. Let the center-of-mass position of the atom be denoted
$\vb r$, the atomic mass $M$, and label the manifold of internal
energy levels that are coupled resonantly by the lasers $\ket{\alpha}$.
We will consider this manifold to be a subset of the ground-state
manifold of the atom, and ignore finite lifetime effects. Upon eliminating the photon dynamics by going
into a rotating frame, the dynamics for the atom are governed by
the familiar sum of kinetic and potential energy $\hat{V}=\sum_{\alpha\beta}V^{\alpha}_\beta(\vb r)\ket{\alpha}\bra{\beta}$ of a spinor particle
\begin{equation} 
\hat{H}=-\frac{\nabla_{\vb r}^2}{2M}
	\hat{1}+\hat{V}(\vb r)\label{eq:ham_basic}\,.
\end{equation} 
Despite atomic neutrality, this Hamiltonian can
give rise to a lowest band similar to a Landau level~\cite{DalibardRMP2011, Goldman_2014,CooperRMP2019}. The key to
understanding this is the spinor nature of the particle, and it is
particularly clear in the adiabatic deep-lattice limit $M\to \infty$.
In this limit, the Hamiltonian reduces to the potential
energy $\hat{V}(\vb r)$. The low-energy manifold consists of
the ground state of $V^\alpha_\beta(\vb r)$ for each $\vb r$, which
forms a spinor texture $n^\alpha(\vb r)$. Consider now including the
leading $1/M$ corrections to the dynamics, which amounts to projecting
\eqref{eq:ham_basic} to the subspace of adiabatic wavefunctions of the
form $\braket{\vb r,\alpha|\psi}=\psi^\alpha(\vb r)=b(\vb r)n^\alpha(\vb r)$ where $b(\vb r)$ is
an arbitrary scalar wavefunction. Since $n^\alpha(\vb r)$ is fixed,
the dynamics are purely those of the scalar $b(\vb r)$. Under an infinitesimal translation the
total wavefunction picks up a Berry phase\cite{QuantalBerry1984} from the change in the spinor direction. This manifests as an emergent background gauge
field acting on $b(\vb r)$, which is evident from projecting the
momentum to the space of adiabatic wavefunctions
\begin{equation} -\i
	\bar{n}_{\alpha}(\vb r)\nabla_{\vb r}\psi^\alpha(\vb r)=[-\i\nabla_\vb
	r-\i\bar{n}_\alpha(\vb r)\nabla_{\vb r}n^\alpha(\vb r)]b(\vb r)\,.
\end{equation} 
Clearly this is the same form as a scalar particle
in the presence of a magnetic field with vector potential 
$\vb A_\mr{texture}=\i\bar{n}_{\alpha}(\vb r)\nabla_{\vb r}n^\alpha(\vb r)$, or equivalently magnetic field 
\begin{equation} 
B_\mr{texture}(\vb r)
=\i\nabla_{\vb r}\bar{n}_\alpha(\vb r)\wedge\nabla_{\vb r}n^\alpha(\vb r) \,.\label{eq:btext} 
\end{equation} 
The Berry phase of the spinor texture becomes the Aharonov-Bohm phase\cite{AB_phase} of the effective magnetic
field felt by $b(\vb r)$\cite{cooper2011optical,DalibardRMP2011,YePRL1999BPT}, and the astute
reader will recognize equation \eqref{eq:btext} as the expression
for the Berry curvature of $n^\alpha(\vb r)$. The Hamiltonian
governing $b$ in the adiabatic limit takes the form
\begin{equation}
	H^\mr{adi}
    =\frac{(-\i\nabla_{\vb r}-\vb A_\mr{texture})^2}{2M}
    + \frac{D(\vb r)}{2M}+\bar{n}_\alpha V^\alpha_\beta n^\beta
    \label{eq:ham_adiabatic} 
\end{equation}
where $D(\vb r)=\nabla \bar{n}_{\alpha}(\delta^\alpha_\beta-n^\alpha\bar{n}_\beta)\nabla n^\beta$
is the trace of the Fubini-Study metric of the
texture. This adiabatic Hamiltonian is valid whenever the
low-energy dynamics can be projected along a spin texture. This is
evidently justified in deep lattices, but even when $V^\alpha_\beta$
is comparable to the typical kinetic energy, there are a plethora
of examples where this holds approximately. 

An optical flux lattice (OFL) is defined to be an optical lattice whose low-energy
subspace has a texture $n^\alpha(\vb r)$  with effective magnetic field
$B_\mr{texture}$ of nonzero average flux.

Perhaps the simplest example -- for a two-level system -- is the triangular OFL introduced in Ref.~\cite{cooper2011optical}, for which 
\begin{equation} 
    \hat{V}_{\rm tri}(\vb r)
    = \sum_{i=1}^3 \hat{\sigma}_i \cos(\vb q_i\cdot \vb r)\,,\label{eq:tri}
\end{equation}
where $\hat{\sigma}_{i=1,2,3}$ are the Pauli matrices, and 
$\vb q_i$ are three in-plane momentum at $120^\circ$ to each other. We define them as  
\begin{equation}
\vb q_i \equiv q\vb e_{ x}\cos\tfrac{2\pi i}{3}+q\vb e_y\sin\tfrac{2\pi i}{3}
 \label{eq:directions}
\end{equation}
in terms of the orthogonal unit vectors $\vb e_{x,y}$ and the overall momentum scale $q$.  
It is readily found that $V(\vb r)$ is gapped at all positions, and the dressed state wraps the Bloch sphere to give nonzero flux of $B(\vb r)$ through the periodic real-space cell. The real-space lattice is triangular, with a high degree of translational symmetry such that $B(\vb r)$ and $D(\vb r)$ have translational symmetry over a cell with just $\pi$-flux of $B(\vb r)$. (This is discussed further in Sec.~\ref{subsec:translational}.)

\begin{figure} \centering
	\includegraphics[width=\linewidth]{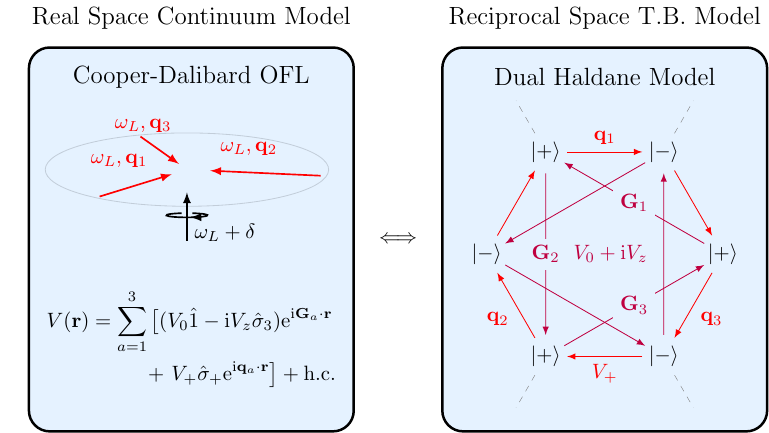}
	\caption{Illustration of how the two-photon optical flux lattice introduced in Ref.~\cite{cooperOpticalFluxLattices2011} can be viewed as a tight-binding model in reciprocal space, which is equivalent to the Haldane model. Interspecies coupling involves one photon at $\omega_L$ and one at $\omega_L+\delta$, with momentum transfers $\vb q_i$. Intraspecies couplings involve two photons at $\omega_L$, with momentum transfers $\vb G_i = \vb q_{i+1} - \vb q_{i}$.}
	\label{fig:reciprocal_spaceillustration} \end{figure}

Practical implementations for OFLs in cold atomic gases are likely to employ two-photon Raman coupling. A simple form of OFL using two-photon coupling was proposed in Ref.~\cite{cooperOpticalFluxLattices2011}. This uses three in-plane beams at frequency $\omega_L$ along directions $\vb q_i$ (\ref{eq:directions}) to form Rabi couplings,
\begin{eqnarray}
\kappa_0 & = & \kappa \cos\theta \sum_{i=1}^3 \e^{\i \vb q_i \cdot \vb r} \,,\\
\kappa_\pm & = & \kappa \sin\theta \sum_{i=1}^3 \e^{\i \vb q_i \cdot \vb r  +\i \hat{\vb e}_z\times \hat{\vb q}_i}\label{eq:kappa-pm}\,,
\end{eqnarray}
where $\kappa$ and $\theta$ are parameters.  The transition between internal states involves one out-of-plane circularly polarized beam at frequency $\omega_L+\delta$ with spatially uniform coupling ${\kappa}'_-$. [See Fig. \ref{fig:reciprocal_spaceillustration}(a).] From these, the potential is of the form
\begin{eqnarray}
\nonumber
\hat{V}_{\rm 2-photon}(\vb r)  & = &     \frac{\kappa_{\rm tot}^2}{3\delta} \hat{1} + 
    \frac{1}{3\delta} \left[(|\kappa_+|^2 - |\kappa_-|^2)\hat{\sigma}_3 \right.
    \\
     &  & \left. + \bar{\kappa}'_-\kappa_+\hat{\sigma}_+ +{  \kappa}'_- \bar{\kappa}_+ \hat{\sigma}_-\right]\,,
     \label{eq:2-photon}
\end{eqnarray} 
where $\kappa_{\rm tot}^2 = \sum_m|\kappa_m|^2$, and $\delta$ is the detuning between the two internal states. See Appendix \ref{app:pol} for details on the polarizability.
This again leads to a real-space unit cell that is triangular, in which the Bloch vector wraps the sphere, signaling that there is a net flux per unit cell. However, in this case the minimal translational symmetry encloses $2\pi$ flux per unit cell.

\subsection{Designing in reciprocal space}

\label{subsec:reciprocal}

The definition of OFLs, and the above examples, were motivated by the properties in real space. However, given that optical lattices involve the use of lasers to write in momentum transfers (with or without changes in internal state), it is very instructive to view them in reciprocal space.
Indeed, this set of momentum transfers directly constructs a tight-binding model in reciprocal space. 
In the adiabatic limit, the effective magnetic field is nothing but
the Berry curvature of $n^\alpha(\vb r)$ which is the ground state of
this reciprocal-space tight-binding model\cite{cooper2012designing}. Hence considering any tight-binding
Chern insulator with Hamiltonian $V^{\alpha}_{\beta}(\vb k)$, there is
a dual optical flux lattice with potential $V^\alpha_\beta(\vb r)$.

To illustrate the reciprocal space mapping, consider the triangular OFL constructed for spin-$1/2$
fermionic atoms in \cite{cooperOpticalFluxLattices2011}, and defined above, Eq.~(\ref{eq:2-photon}). 
The interspecies coupling involves one in-plane photon, from $\kappa_{\pm}$, and one out-of-plane photon from $\tilde{\kappa}$. Thus, from (\ref{eq:kappa-pm}), it involves momentum transfers of the three $\vb q_i$. These couplings construct the nearest-neighbor bonds on a honeycomb lattice in reciprocal space, with $|\pm\rangle$ at the two inequivalent sites. The intraspecies couplings involve two in-plane photons, thus giving momentum transfers of  $\vb G_2 = \vb q_3-\vb q_2$ etc. These couplings give the  next-nearest-neighbor bonds on the honeycomb lattice.
Overall the optical potential of the two-photon OFL of  Ref.~\cite{cooperOpticalFluxLattices2011} generates precisely the Haldane model in reciprocal space. See Fig.~\ref{fig:reciprocal_spaceillustration}(b). This tight-binding model is in its topological phase, with unit Chern number,  consistent with the fact that the corresponding OFL has $2\pi$ flux per real space unit cell\cite{cooperOpticalFluxLattices2011}.

This is one of the model potentials that we shall focus on in the rest of the paper. Although it was first introduced in Ref.~\cite{cooperOpticalFluxLattices2011} as a model for spin-1/2 atoms in an optical flux lattice, in light of the connections highlighted above, we shall refer to it as the {\it dual Haldane model}, and denote the potential as $\hat{V}^{\hexagon}$ to represent the hexagonal lattice in reciprocal space.
For generality, we shall add a scalar potential, and write the model potential as \begin{equation} \hat{V}^{\hexagon}
	(\vb r)=\sum_{a=1}^3 (V_0\hat{1}-\i V_z\hat{\sigma}_3)\e^{\i\vb G_a\cdot \vb r
	}+V_+\hat{\sigma}_+\e^{\i\vb q_a\cdot \vb r}+\mathrm{h.c.} \,, \label{eq:pot_hal}
\end{equation}  with parameters
$V_{0,z,+}$. 

Note that the mapping to the dual Haldane model applies only to the optical potential $\hat{V}({\vb r})$ and therefore to the real-space spinor texture; the model of Ref.~\cite{cooperOpticalFluxLattices2011} also includes the kinetic energy term of (\ref{eq:ham_basic}) so its energy bands depend on the ratio of kinetic to potential energies. The kinetic energy is quadratic in momenta, and so the total single-particle energy is obtained by introducing a momentum space `harmonic oscillator potential', whose minimum is located at the location of the  crystal momentum under consideration $\vb k$.
{This same model is also an effective Hamiltonian for electrons in a single valley of twisted transition metal dichalcogenide homobilayers\cite{wu2019topological,MoralesPRL2024MATTMD}. To understand why, it is helpful to think in reciprocal space, as above: the OFL of Cooper and Dalibard~\cite{cooperOpticalFluxLattices2011}  is precisely the Haldane model written in reciprocal space -- the dual Haldane model considered here -- which is the simplest two-state tight-binding model with triangular symmetry and short-range hopping. It is therefore unsurprising that the simplest two-state model describing a single TMD valley, sharing the same symmetry and minimizing the number of wavevector transfers, coincides with it exactly. This identification of the two models shows the power of designing Chern bands in reciprocal space.} In moiré materials the overall ratio of kinetic to potential energy is controllable by the twist angle, though the ratio of the potential parameters is determined chemically, and may also depend on the twist angle through lattice relaxation. As each potential parameter is individually tunable in cold atomic gases, the dual Haldane model is far more flexible in this setting, and furthermore comparatively easy to characterize. In particular, in the general model (\ref{eq:pot_hal}) we have added the freedom to tune the scalar potential $V_0$ separately from the other couplings.

\begin{figure}
    \centering
    \includegraphics[width=\linewidth]{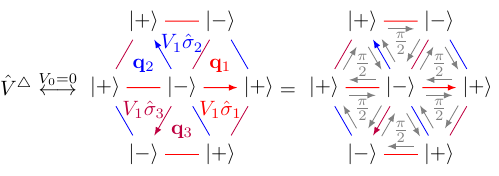}
    \caption{The momentum space couplings of $\hat{V}^\triangle$ in the absence of the scalar potential form a nearest-neighbor $\pi/2$ flux triangular lattice.}
        \label{fig:vtriangle}
\end{figure}

Similarly,  we adapt the potential of the triangular model (\ref{eq:tri})  to include a scalar potential $V_{0}$, beyond the case $V_0=0$ 
considered in \cite{cooper2011optical}. 
We call this the \textit{dual triangular model} -- strictly it is a nearest-neighbor triangular lattice in reciprocal space at $\pi/2$ flux (see Fig.~\ref{fig:vtriangle}). We
write it in the form \begin{equation}
\hat{V}^\triangle(\vb r)=\sum_{a=1}^3V_1
	\hat{\sigma}_a\e^{\i\vb q_a\cdot \vb r}+ V_0\hat{1}\e^{2\i\vb q_a\cdot \vb
		r}+\text{h.c.}
        \,.
        \label{eq:pot_tri} \end{equation}
Finally, a square version of this model, the \textit{dual square model}, can be
constructed similarly using next-nearest-neighbor momentum transfers.
Let $\tilde{\vb q}_{1,2,3,4}$ be $\pi/2$-rotated momenta of equal
magnitude, $\tilde{\vb G}_1=\tilde{\vb q}_{2}-\tilde{\vb q}_1$
along with cyclic permutations, and $\hat{\sigma}_a^\perp=\hat{\sigma}_{a\text{
			mod 2}}$. The potential we consider adds a scalar lattice
to the model in \cite{cooper2011optical,juzeliunas2012flux}
and takes the form: \begin{align} \hat{V}^\square(\vb r) &
              =\sum_{a=1}^4 V_+\hat{\sigma}_{a}^\perp\e^{\i\tilde{ \vb q}_a\cdot\vb
	              r}+V_z\hat{\sigma}_3\e^{\i\tilde{\vb G}_a\cdot \vb r}+V_0\hat{1}\e^{2\i \tilde{\vb
		              G}_a\cdot \vb r}\label{eq:pot_square}\end{align} where $V_{0,z,+}$
are tunable potential strengths. Strictly this dual square lattice is a next-nearest-neighbor square lattice in reciprocal space at $\pi/2$ flux per triangular plaquette.

\subsection{Translational Symmetries}
\label{subsec:translational}

We shall make much use of symmetries in later sections of the paper.
It is therefore important to pause to consider in detail the symmetries of the three lattices just presented.

Arising from standing beams of light in a
periodic interference pattern, the potential is genuinely periodic.
Nevertheless, a finer translational symmetry can often be found by
letting the translation operators act on the spinor degrees of freedom.

For the dual triangular (square) model, the minimal unit cell under which
$\hat{V}^\triangle$ ($\hat{V}^\square$) is periodic has
reciprocal lattice $\Lambda^\star_\mr{pot}=2\vb q_1\Z\oplus2\vb q_2\Z$
($\Lambda^\star_\mr{pot}=2 \tilde{\vb q}_1\Z\oplus2\tilde{\vb
		q}_2\Z$). Let $\vb R_{1,2}$ be the dual basis of
$\Lambda_\mr{pot}$, then the translation operators $\hat{T}_{1}=\e^{\vb
		R_1\cdot\vb \nabla}\hat{\sigma}_2,\hat{T}_2=\e^{\vb R_2\cdot\vb \nabla}\hat{\sigma}_1$
are symmetries $\hat{T}_a\hat{H}=\hat{H}\hat{T}_a$. However, they only furnish a projective
representation of the translation group, as they do not mutually commute
$\hat{T}_1\hat{T}_2=-\hat{T}_2\hat{T}_1$. This is because the \textit{real space} unit cell of $\Lambda_\mr{pot}$
only encloses $\pi$ flux. By doubling the unit cell, a
commuting translation algebra is generated by $\hat{T}^2_1,\hat{T}_2$\cite{cooper2011optical}. To compute a band structure we therefore use the lattice $\Lambda=2\vb R_1\Z\oplus\vb R_2\Z$. Nevertheless, the finer true symmetry makes its existence felt by restricting the possible non-vanishing Fourier component of observables. Noting the \emph{real space} flux and 6-fold $C_6$ (4-fold $C_4$) rotational symmetry of $\Lambda^\star$, we refer to this situation as $C_6$ $\pi$-flux  ($C_4$ $\pi$-flux) symmetry.

For the dual Haldane model, $\hat{V}^{\hexagon}$, the finest translation symmetry is commuting and has $C_6$ symmetric reciprocal
lattice $\Lambda^\star=\Lambda^\star_\mr{pot}=\vb G_1\Z\oplus\vb G_2\Z$.
We call this $C_6$ $2\pi$-flux symmetry, and the analogous situation for a square lattice would be $C_4$ $2\pi$-flux symmetry.

\section{Ideality}
\label{sec:ideal}
Having discussed several examples of OFLs, let us return to the adiabatic Hamiltonian \eqref{eq:ham_adiabatic}. A flat Chern band is evidently furnished from the Landau levels when the magnetic field and potential $\frac{D}{2M}+\bar{n}_\alpha V^\alpha_\beta n^\beta$ are  both uniform. Perhaps surprisingly, as was recognized in \cite{aharonovGroundStateSpin1/21979}, and later in the context of adiabatic moiré bands\cite{MoralesPRL2024MATTMD,shiAdiabaticApproximationAharonovCasher2024}, a flat lowest band can survive a non-uniform magnetic field $B_\mr{texture}(\vb r)$ with the same average, when the potential takes the value $\frac{D}{2M}+\bar{n}_\alpha V^\alpha_\beta n^\beta=-\frac{B_\mr{texture}}{2M}$. The Hamiltonian then reduces to the Aharonov-Casher (AC) form:
\begin{equation}
    H^\mr{AC}=\frac{[-\i\nabla_{\vb r}-\vb{A}_\mr{texture}(\vb r)]^2}{2M}-\frac{B_\mr{texture}(\vb r)}{2M}\,.\label{eq:ham_ac}
\end{equation}
To see how this occurs, let us introduce the following notation to draw similarities with the Landau levels. (A more complete review is given in Appendix \ref{app:landau}.) Denote the unit cell area of $\Lambda$ containing $2\pi$ flux by $2\pi \ell_B^2$, and
the associated cyclotron frequency $\omega_c=(M\lB^2)^{-1}$. We use the complexified coordinate $z=x+\i y$, while the
complexified kinetic momentum for the corresponding uniform field (in symmetric gauge) is
$\pi=-\i (\partial_x-\i\partial_y)+\i \bar{z}/2\ell_B^2$. We will also need the
 momenta of opposite chirality, which can be considered the `guiding-center' momenta, $\tilde{\pi}=\pi-\i \bar{z}/\ell_B^2$. A non-uniform field with average $\lB^{-2}$ can be described by the complexified vector potential $\mc{A}=\i\bar{n}_\alpha(\partial_x+\i\partial_y)n^\alpha=\i{z}/2\lB^2+\delta \mc{A}$. Thus the AC Hamiltonian can be rewritten as 
\begin{equation}
    H^\mr{AC}=\frac{(\pi-\delta\bar{\mc{A}})(\pi^\dagger-\delta \mc{A})}{2M}\,.
\end{equation}
This Hamiltonian naturally has zero modes equal in quantity to the number of solutions to 
\begin{equation}
    [\pi^\dagger-\delta\mc{A}]b(\vb r)=0.\label{eq:zeromode}
\end{equation}
When $\delta\mc{A}=0$, this is nothing but the usual lowest Landau level (LLL) condition, with solutions for each crystal momentum $\vb k$ in the Brillouin zone of $\Lambda^\star$ given by
\begin{equation}
	\braket{\vb r|\vb k}=\mc{N}\sigma(z-\i \ell_B^2 k)\e^{-(\ell_B^{-2}|z|^2+\ell_B^2 |k|^2)/4+\i \bar{k} z/2},
\end{equation}
where $\sigma$ is the modified Weierstrass $\sigma$-function\cite{haldane2018modular}, $k=k_x+\i k_y$, and $\mc{N}$ is a normalization constant. Furthermore, whenever $\delta\mc{A}$ corresponds to a magnetic field modulation $\delta B_\mr{texture}=B_\mr{texture}-\lB^{-2}$ with vanishing average, the generic solution to \eqref{eq:zeromode} takes the form of density modulated lowest Landau levels
\begin{equation}
    b_\vb k(\vb r)=\e^{\phi(\vb r)}\braket{\vb r|\vb k};\qquad \pi^\dagger\phi=\delta \mc{A}\label{eq:AC_sols}\,.
\end{equation}
In the Coulomb gauge $\nabla_{\vb r}\cdot \vb A_\mr{texture}=0$ the density modulation $\e^{\phi}$ is the K\"ahler potential for the magnetic field modulation
\begin{equation}
   - \nabla^2\phi=\delta B_\mr{texture}(\vb r)\,.
\end{equation}
The manifold of zero modes forms a flat band with Chern number $\mc{C}=-1$. Beyond being a flat Chern band, the LLL is remarkable for the ability to attach vortices within it, so that $b(\vb r)\mapsto zb(\vb r)$ does not cause any excitation. Chern bands for which vortex attachment is possible are known as ideal or vortexable\cite{ledwith2023vortexability,wang2021exact}. Wavefunctions belonging to many fractional quantum Hall phases in the LLL have exact parent Hamiltonian given by various zero-range repulsive interactions, with respect to which they have zero energy. For such parent interactions, ideal bands that are flat also give rise to zero energy ground states in the same phase as the LLL. 
For scalar particles with $\mc{C}=-1$ (and $2\pi$ flux per unit cell), the AC wavefunctions \eqref{eq:AC_sols} exhaust the possibilities for ideal wavefunctions\cite{estienne2023ideal}. 

Instead of approaching the design of OFLs with a view towards mimicking a uniform field, a natural path to flat ideal bands is AC Hamiltonian mimicry. For the general case on which we first focus, we will not be able to find an exact AC Hamiltonian. Nevertheless, we shall show that perturbations to an AC Hamiltonian are well behaved, even when they are comparable in size to the typical gap $\sim\omega_c$.
Including an amplitude modulated spinor $\chi^\alpha(\vb r)=n^\alpha (\vb r)\e^{\phi(\vb r)}$, a $\mc{C}=-1$ ideal band in an OFL has wavefunction
\begin{equation}
    \psi^\alpha_{\vb k}(\vb r)=\chi^\alpha(\vb r)\braket{\vb r|\vb k}\label{eq:ideal_def}\,.
\end{equation}
A useful practical criterion for ideality is saturating the trace inequality, which we now describe.
To any wavefunction with crystal momentum $\vb k$ there is a periodic
Bloch wavefunction of the form
$\braket{\vb r\alpha|{u_{\vb k}}}=\e^{-\i\vb k\cdot\vb r}\psi^\alpha_{\vb k}(\vb r)$. From this we
may compute the Berry curvature $\Omega(\vb k)=\i \braket{\nabla_{\vb
		k}u_{\vb k}|\wedge|\nabla_{\vb k}u_{\vb k}}$ and Fubini-Study metric
$g^{ab}(\vb k)=\braket{\nabla^a_{\vb k}u_{\vb k}|(1-\ket{u_{\vb
				k}}\bra{u_{\vb k}})|\nabla^b_{\vb k}u_{\vb k}}$. For any band
$\tr g(\vb k)\geq|\Omega(\vb k)|$ with equality for all $\vb
	k$ only if the band is ideal\cite{wang2021exact}. Noting that
$\int\dd^2\vb k\,\Omega(\vb k)=2\pi\mc{C}$, a natural measure of
deviation from ideality is the trace violation 
\begin{equation}
	T=\left[\frac{1}{2\pi}\int \dd^2\vb k\tr g(\vb k)
		\right]-|\mc{C}|\geq0\label{eq:trace_violation} \,,
\end{equation} 
which takes the value $2n$ for the $n$th Landau level\cite{TomokiPRB2021TraceCondnLLL}, see Appendix \ref{app:landau_trace} for a derivation. For comparison with a different measure of ideality deviation used in \cite{nascimbene2024emergence} see Appendix \ref{app:circular_dichroism}.

\subsection{{Form factor filtering}}
With the preparation of the previous sections, let us wrest nearly flat and ideal bands from the flexibility
afforded by the optical lattice potential. The key
principle that will permit the near-flat ideal Chern bands  in  general cases is that the size
of the cyclotron orbits significantly suppresses small-scale variations,
as measured by the magnetic form factor
\begin{align}
	I_{\vb G\in\Lambda^\star}(\vb k) & =\braket{\vb k|\e^{\i\vb G\cdot\vb r}|\vb k}                                                             \\
	                                 & =\braket{\vb k|\e^{\ell_B^2G(\pi-\tilde{\pi})/2 +\ell_B^2 \bar{G}(\tilde{\pi}^\dagger-\pi^\dagger)/2}|\vb k} \\
	                                 & \label{eq:formfactor}=\e^{\lB^2[\i\vb G\wedge \vb k-\vb G^2/4]}\begin{cases}
		                                                                             +1 & \vb G/2\in\Lambda^\star \\
		                                                                             -1 & \text{otherwise}
	                                                                             \end{cases}
\end{align}
{This suppression acts as a strong low-momentum filter on many quantities of interest. The strength of this filter is controlled by the symmetry allowed reciprocal lattice vectors, which can naturally be ordered into shells of increasing momentum magnitude. For the models which we seek to study, we note the relevant magnitudes in Table \ref{table:shellformfactor}.}
\begin{table}[h]
\centering
	\includegraphics{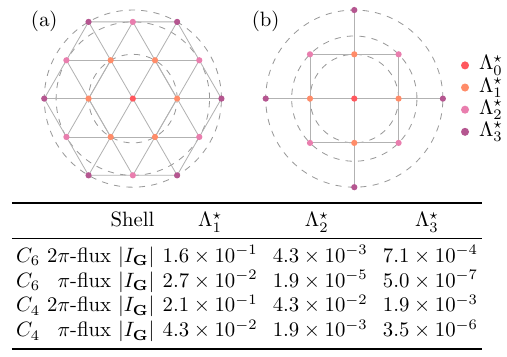}
	\caption{Illustration of momentum shells for (a) $C_6$-symmetric and (b) $C_4$-symmetric lattices. Form factors \eqref{eq:formfactor} for the inner three shells are tabulated for models with $2\pi$ and $\pi$-flux symmetries.}
	\label{table:shellformfactor}
\end{table}
\subsection{{Ideal perturbation theory}}
{To go beyond the Aharonov-Casher limit of  ideal wavefunctions as zero modes of \eqref{eq:ham_ac}, a non-unitary basis change is illuminating. Instead of working with the typical Bloch basis,  we factorize a LLL piece, so that a generic wavefunction $\psi_\vb k^\alpha(\vb r)$ is specified by the spinor $\chi_{\vb k}^\alpha(\vb r)$ as
$\psi^\alpha_{\vb k}(\vb r)=\chi_{\vb k}^\alpha(\vb r)\braket{\vb r|\vb k}$. In this basis, the space of ideal wavefunctions are those where $\chi_{\vb k}$ is momentum independent.}
Consider a situation where the ideal wavefunction \eqref{eq:ideal_def} describes the lowest band. It is generically
dispersive, as can be computed by the Rayleigh quotient
\begin{align}
	E_{\vb k}[\chi^\alpha] & =
	\frac{\int \dd \vb r~|\langle{\vb r|\vb k}\rangle|^2\left[\frac{1}{2M}|\tilde{\pi}^\dagger\chi^\alpha|^2+\bar{\chi}_\alpha V^\alpha_\beta \chi^\beta\right]}
	{\int \dd \vb r~|\langle{\vb r|\vb
	k}\rangle|^2\bar{\chi}_{\alpha}\chi^\alpha}  \\
	                       & =\frac{\sum_{\vb G}I_{\vb G}(\vb k)\left[\frac{1}{2M}|\tilde{\pi}^\dagger\chi^\alpha|^2+\bar{\chi}_\alpha V^\alpha_\beta \chi^\beta\right]_{\vb G}}{\sum_{\vb G}I_{\vb G}(\vb
	k)[\bar{\chi}_\alpha\chi^\alpha]_{\vb
			G}}\label{eq:dispersion}
\end{align}
where the notation $[\cdot]_{\vb G}$ is shorthand for the $\vb G$ Fourier
component of the bracketed quantity. 
The Gaussian decay of $|I_{\vb G}|$ ensures that the band can be made exponentially flat in the number of tuning parameters of $\chi^\alpha$. To see this, imagine tuning the Fourier components of the numerator and denominator
for the $N$ innermost shells $\vb G\in\Lambda_{0,1,2,\dots N}$ to have a common ratio
\begin{equation}
	E_0=\left[\frac{1}{2M}|\tilde{\pi}^\dagger\chi^\alpha|^2+\bar{\chi}_\alpha V^\alpha_\beta \chi^\beta\right]_{\vb G}/[\bar{\chi}_\alpha\chi^\alpha]_{\vb
	G}.
\end{equation}
Then the first term that causes dispersion is suppressed by $|I_{\vb G\in\Lambda_{N+1}^\star}|$ compared to the typical energy scale. As long as the potential is at most comparable to $\omega_c$, we expect a bandwidth comparable to $\omega_c|I_{\vb G\in\Lambda^\star_{N+1}}|$, which decays extremely quickly with $N$ as seen in Table \ref{table:shellformfactor}. Let us refer to this situation as an $N$-flat band, and the manifold of model parameters on which this is achieved, the $N$-flat manifold.
In a similar discussion, focused on the adiabatic limit for the dual Haldane model, finding a $1$-flat manifold was taken as the `magic angle' condition for twisted transition metal dichalcogenides\cite{shiAdiabaticApproximationAharonovCasher2024}.
The power of optical lattices in cold
gases is that one can shape the potential with tuning parameters, so as to reach the $1$- or $2$-flat regimes, or potentially beyond. {Reaching a higher $N$-flat condition requires tuning additional, independent Fourier harmonics of the scalar potential. This does not require additional laser beams: as illustrated for the dual Haldane model in App.~\ref{app:DualHaldaneExp}, the extra harmonics can instead be added as extra frequency tones on the beams already present, for example using an acousto-optic or electro-optic modulator. Since these tones need only contribute to the scalar potential, they can also be far-detuned, and so do not add appreciably to the photon scattering budget.}

This dispersion estimate is only
reliable when the band is similarly close to ideality, otherwise the dispersion
would be dominated by the $\vb k$ dependence of the true $\chi^\alpha$. To understand this dependence, consider the reformulation of the genuine eigenvalue equation in terms of the spinor $\chi_{n\vb k}^\alpha$ which depends on the band index $n$ and the crystal momentum $\vb k$: 
\begin{align}
	E_{n\vb k}\chi^\alpha_{n\vb k}
	 & =\tilde{H}^\alpha_{\beta}\chi^\beta_{n\vb k}+\hat{\zeta}_{\vb
		k}\chi^{\alpha}_{n\vb
	k}\label{eq:exact}                                                                                                                                                                             \\
	 & =\left(\frac{\tilde{\pi}\tilde{\pi}^\dagger}{2M}\delta^\alpha_\beta+V^\alpha_\beta\right)\chi^\beta_{n\vb k}-\frac{\i}{M}\hat{\zeta}(z-\i\lB^2 k)\tilde{\pi}^\dagger\chi^\alpha_{n\vb k}\nonumber
\end{align}
where $\hat{\zeta}(z)=\zeta(z)-\bar{z}/2\lB^2=-\i\sum_{\vb
		G\neq \vb 0}\frac{\e^{\i\vb G\cdot \vb r}}{\lB^2 G}$ is the periodic completion of the
Weierstrass $\zeta$ function\cite{haldane2018modular}. We warn that $\hat{\zeta}$ has poles at the lattice points, arising from the zeros of $|\langle \vb r|\vb k\rangle|^2$. This eigenvalue equation may be readily implemented numerically using the matrix elements computed in Appendix \ref{app:lllform_factor}.
\begin{figure*}[tb]
	\centering

	\includegraphics[width=1.\textwidth]{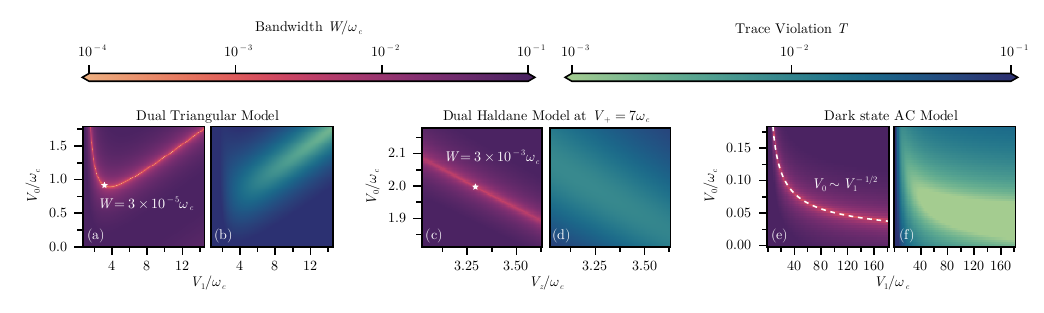}
	\caption{Bandwidth (a,c,e) and trace violation (b,d,f).
    In all three cases clear codimension-$1$ manifolds of extremely narrow bands are visible. The minimal bandwidths of the dual triangular and Haldane models are indicated by $\star$, and match the expectation for the $1$-flat manifold given their $C_6$ $\pi$-flux respectively $2\pi$-flux symmetries. In contrast, the trace violation achieves its minimum close to the $1$-flat manifold in all cases, but grows slowly in the surrounding region (it remains far below the value $T=2$ characteristic of the $1$st Landau level).}
	\label{fig:phase-diag}
\end{figure*}

In an ideal band, the same spinor $\chi^\alpha$ is a solution to equation \eqref{eq:exact}
for all $\vb k$, hence also the $\vb k$ average thereof
\begin{equation}
	E_\star \chi^\alpha=\tilde{H}^\alpha_\beta \chi^\beta \,. \label{eq:idealHam}
\end{equation}
We take the solution of this equation to be our ansatz for the optimal ideal wavefunction away from the AC limit. This is especially reasonable since  $\chi^\alpha$ also minimizes the $\vb k$ averaged energy of the lowest band \eqref{eq:dispersion}, albeit the average with weighting factor $w(\vb k)=\int\dd \vb r~|\langle\vb r|\vb k\rangle|^2 \bar{\chi}_\alpha\chi^\alpha$. The uniformity of this weighting factor is directly controlled by the Berry curvature distribution $\Omega({\vb k})=-\ell_B^2+\nabla^2_{\vb k}\log w(\vb k)$\cite{wang2021exact}, being the K\"ahler potential for its modulation.
{The benefit of the aforementioned perspective, is that the perturbation theory about ideality is well controlled. Let $U$ be the characteristic size of the energetic perturbation from ideality $\braket{\vb k|\bar{\chi}_{m\vb k}\hat{\zeta}\chi_{n\vb k}|\vb k}$; elementary perturbation theory then determines the resulting admixture of excited bands into the ideal wavefunction, order by order in $U/\omega_c$.}

Expanding the ideal wavefunction in the actual eigenbasis $\chi^\alpha=\sum_{n=0}^\infty c_n(\vb k)\chi_{n\vb
		k}^\alpha $ we may estimate the overlap with the lowest band $c_0(\vb k)$ by neglecting the mixing between
excited states. This amounts to letting $c_n(\vb k)=c_0(\vb k)
	\frac{\braket{\vb
			k|\bar{\chi}_{n\vb k} \tilde{H}\chi_{0\vb k}|\vb k}}{E_\star -
		\braket{\vb k|\bar{\chi}_{n\vb k}\tilde{H}\chi_{n\vb k}|\vb k}}$. The denominator
is essentially the difference in energy between the $0$th and $n$th band, which generically grows linearly in $n$. The
numerator by contrast is strongly suppressed by the form factor since
\begin{align}
	\braket{\vb
	k|\bar{\chi}_{n\vb k} \tilde{H}\chi_{0\vb k}|\vb k} & =-\braket{\vb k|\bar{\chi}_{n\vb k}\hat{\zeta}_{\vb k}\chi_{0\vb k}|\vb k}
	\\&=\frac{\i}{M}\sum_{\vb G\in \Lambda}\bar{G} I_{\vb G}(\vb k)[\bar{\chi}_{n\vb k}\tilde{\pi}^\dagger \chi_{0\vb k}]_{\vb G}
\end{align}
Recall that the quantity $[\bar{\chi}_{n\vb k}\tilde{\pi}^\dagger \chi_{0\vb
				k}]_{\vb G}$ vanishes in the AC limit. If $U$ is the characteristic
energy scale associated with the deviation therefrom, equation \eqref{eq:idealHam} implies that $\chi$ is modified at order $U/\omega_c$. Neglecting numerical constants except for the Gaussian we generically expect
$|c_0(\vb k)|^2 /w(\vb k)\sim 1- \e^{-\lB^2\vb G_1^2/2} U^2/\omega_c^2$ where $\vb G_1\in\Lambda^\star_1$. By contrast the overlap
between the AC and optimal $\chi$ is $\sim 1-U^2/\omega_c^2$.
This suggests that the shape of the ideal
wavefunction is easily perturbed, but near-ideality (overlap with an ideal wavefunction comparable to $1-|I_{\vb G\in\Lambda_1^\star}|^2$) is considerably more robust. Taking this effect into account, the true lowest band dispersion $E_{0\vb k}$ is
\begin{equation}
    E_{0\vb k}=E_{\vb k}+\sum_{n>0}\frac{(E_\vb k-E_{n\vb k})|\braket{\vb k|\bar{\chi}_{n\vb k}\hat{\zeta}_{\vb k}\chi_{0\vb k}|\vb k}|^2}{(E_\star-\braket{\vb k|\bar{\chi}_{n\vb k}\tilde{H}\chi_{n\vb k}|\vb k})^2}
\end{equation}
The difference from the ideal estimate is suppressed to have order of magnitude $U^2|I_{\vb G\in\Lambda^\star_1}|^2/\omega_c$. Even when $U\sim\omega_c$ this is comparable to the $1$-flat dispersion, while the quadratic $U$ dependence hints that sufficiently close to an AC limit, the $2$-flat, and beyond, manifold may be achievable. 

Consider an optical lattice model with tunable potential parameters $V_0,V_1,\dots$
controlling the lattice potential at various shells.
If there is some parameter point $V_0^\star,V_1^\star,\dots$ for which the model forms a nearly ideal OFL, our heuristics indicate the following
features
\begin{enumerate}
	\item The model is nearly ideal in a parameter region of extent comparable to
	      $\omega_c$ surrounding $V_0^\star,V_1^\star,\dots$.
	\item If a model possesses the same rotational symmetry as the lattice $\Lambda^\star$, a codimension-$N$ submanifold is $N$-flat, though for $N>1$ this may only be resolvable sufficiently close to an AC limit.
	\item Since it is possible to reach $N$-flatness by approaching the AC limit for any spinor
	      texture, we expect that $N$-flatness can be achieved
	      by adding only additional scalar lattice parameters to an OFL.
\end{enumerate}
\section{Example Ideal Optical Flux Lattices}
\label{sec:examples}

\begin{figure}
	\centering
	\includegraphics[width=0.9\linewidth]{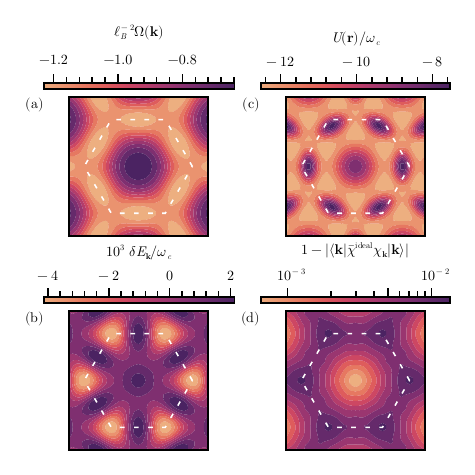}
	\caption{Characteristics of the dual Haldane model at parameters $(V_z,V_+,V_0)=(3.06,7.06,2.09
)\omega_c$. The Berry curvature (a) fluctuates $\Omega(\vb k)=-\lB^2(1\pm0.2)$. By contrast the dispersion (b), which is measured against the average of the lowest band $E=-11.25\omega_c$, is nearly flat. Further, the overlap (d) between the true wavefunction, as computed by \eqref{eq:exact}, and the optimal ideal wavefunction solving \eqref{eq:idealHam} is nearly unity across the entire Brillouin zone. The residual potential deviation from the adiabatic AC limit (c) fluctuates in the range $U(\vb r)\in [-12.5,-7.5]\omega_c$. }

	\label{fig:dualhaldane}
\end{figure}

Our heuristics provide a lens through which to search the parameter space of the models we constructed in Sec. \ref{subsec:reciprocal}. Consider first the dual triangular model \eqref{eq:pot_tri} and dual Haldane model \eqref{eq:pot_hal}. In Fig. \ref{fig:phase-diag} we take two-dimensional parameter cuts, in the latter case fixing $V_+=7\omega_c$, and calculate their bandwidths and the trace violations \eqref{eq:trace_violation}. In sharp agreement with our considerations, we see a parameter region of size $\sim\omega_c$ where the trace violation is comparable to $10^{-2}$ or less in both models. 
Intersections of the centers of these parameter regions are clear codimension-$1$ flat manifolds, with minimal bandwidths of $3.4\times10^{-5}\omega_c$ in the dual triangular case or $3.5\times10^{-3}\omega_c$ in the dual Haldane case. This matches the expectation for $1$-flat bands from Table \ref{table:shellformfactor}. While it may be tempting to believe that these ultra-narrow bands arise because the models are similarly close to having an AC form, that belief is not true. Consider for example the parameter point $(V_z,V_+,V_0)=(3.06,7.06,2.09
)\omega_c$ of the dual Haldane model which lies along the $1$-flat manifold. We plot various characteristics of this parameter point in Fig. \ref{fig:dualhaldane}. The band is essentially ideal, as can be measured by the trace violation of $0.009$, and the overlap between the true and ideal wavefunction predicted by $\tilde{H}$ is $|\braket{\vb k|\bar{\chi}\chi_{0\vb k}|\vb k}|=0.995\pm0.002$ across the Brillouin zone, with the quoted variation being the standard deviation. Surprisingly we note that directly optimizing the average square overlap between the true lowest band and the ideal wavefunction via gradient descent, similar to \cite{li2025variational}, yields an average overlap only $10^{-4}$ better than the ground state of $\tilde{H}$. The bandwidth is $6\times 10^{-3}\omega_c$, and the band gap is $0.7\omega_c$ in agreement with the $1$-flat condition for models with $C_6$ $2\pi$-flux symmetry. In contrast to the uniform Landau level, there is no expectation that the Berry curvature is uniform, or that the excited bands are flat. The former fluctuates $-\lB^2(1\pm0.2)$, and the bandwidth of the first excited band is $0.7\omega_c$. 
To understand the proximity to the AC limit, we compute the potential difference between the AC and adiabatic Hamiltonian for the ground state texture at this parameter point
\begin{equation}
    U(\vb r)=\frac{B_\mr{texture}(\vb r)+D(\vb r)}{2M}+\bar{n}_\alpha V^\alpha_\beta n^\beta\,.
\end{equation}
The root mean square variation of $U(\vb r)$ is $1.2\omega_c$ across the unit cell, with a range of $5\omega_c$. Naively, it is immensely surprising, and highlights the power of our heuristics, that an AC deviation comparable to the single particle gap destroys neither ideality nor the flat band.

In principle the bandwidth of the dual Haldane model could be reduced even more, by tuning to the 2-flat condition. This would be achieved by adding a further scalar potential to the model (\ref{eq:pot_hal}), with wavevectors that lie on the second reciprocal space shell $\Lambda^*_2$. Instead, we illustrate the 2-flat condition for the dual square lattice.

\subsection{2-Flat bands in the dual square model}
\label{subsec:twoflat}
\begin{figure}
	\centering
	\includegraphics[width=\linewidth]{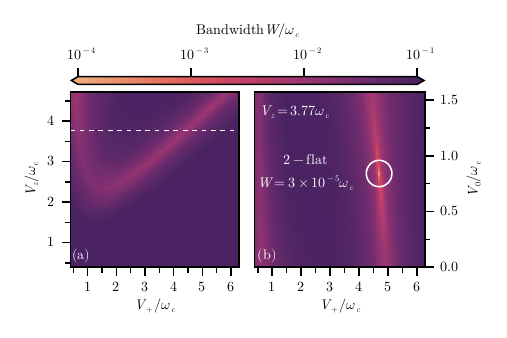}
	\caption{Bandwidth of dual square model with potential \eqref{eq:pot_square} along parameter cuts with (a) $V_0=0$ and (b) $V_z=3.77\omega_c$. A clear $1$-flat manifold of codimension-$1$ can be seen with bandwidth comparable to $10^{-2}\omega_c$, while the circled point indicates the presence of the $2$-flat manifold with bandwidth $3\times 10^{-5}\omega_c$ }
	\label{fig:2flatmanifold}
\end{figure}

Using the three tuning parameters $V_{0,+,z}$ of the dual square model (\ref{eq:pot_square}) we find it possible to reach not just  $1$-flatness, but also the $2$-flat manifold of parameters. Because it is the simplest parameter regime, consider first $V_0=0$. In Fig. \ref{fig:2flatmanifold} we see the clear presence of a $1$-flat manifold, with minimal bandwidth $7\times 10^{-3}\omega_c$. This is in fair agreement with the second form factor $|I_{\vb G\in\Lambda^\star_2}|=2\times10^{-3}$. It is worth emphasizing that the $1$-flat condition  is less restrictive in square lattices than it is in triangular lattices, as the second shell is closer to the first shell in the former case, see the illustration in Table \ref{table:shellformfactor}. To achieve a bandwidth comparable with the $1$-flat triangular model, we instead need to reach the $2$-flat manifold. To do this we turn on the additional scalar lattice $V_0$. Remarkably, we find that for a range of parameter cuts a $2$-flat manifold is apparent. For example taking $V_z=3.77\omega_c$, which is illustrated in Fig. \ref{fig:2flatmanifold}, a codimension-$2$ manifold (in this case a point) of bands with width $3\times 10^{-5}\omega_c$ appears. This is slightly larger than predicted by the $2$-flat condition, which we suspect is due to the non-ideal perturbations.
\subsection{Dark OFL}
\label{sec:darkofl}
From the dual Haldane example it is clear that $1$-flat bands can be achieved by adding a suitably tuned scalar lattice even when the adiabatic Hamiltonian is far from the AC limit. {Flatness achieved this way does not, however, guarantee vortexability: flatness and idealness are a priori independent properties of a band, and a scalar potential tuned only to suppress the bandwidth need not produce an ideal one. We will now introduce an OFL that is an \emph{exact} AC Hamiltonian in the adiabatic limit --- a qualitatively stronger result. While the AC Hamiltonian has been developed as a theoretical ideal~\cite{MoralesPRL2024MATTMD,shiAdiabaticApproximationAharonovCasher2024}, realizations in condensed matter systems achieve this structure only approximately. The dark-state OFL constructed below achieves it exactly, yielding perfectly flat and exactly vortexable Chern bands as an analytical statement. This provides a physical platform in which their idealized limit is realized precisely, and against which the quality of AC approximations in moiré systems can be calibrated.} Nevertheless adding a tunable scalar lattice still permits a huge further decrease in the bandwidth beyond the adiabatic limit.
As was shown in \cite{nascimbene2024emergence}, OFLs can be constructed
also for so-called ``dark" lattices, for which the optically dressed state is exactly decoupled from the optical fields, and lies at the lowest energy with all other ``bright" levels being energetically repulsed. Dark optical lattices can be understood
straightforwardly from their dual tight-binding models. Including both $N$
ground states $\alpha$ and $\tilde{N}$ excited states $\tilde{\alpha}$,
a dark lattice arises when there are no potential couplings connecting
ground states $V^\alpha_\beta(\vb r)=0$, and the number of coupled
ground states exceeds the number of excited states $N>\tilde{N}$. Then
the off-diagonal part of the potential $V^{\tilde{\alpha}}_\beta(\vb
	r)$ must have a kernel of dimension $N-\tilde{N}>0$, which describes textures
with exactly zero potential energy, the ``dark'' states. When the dark-state texture
has net flux per unit cell, these are dark OFLs. It is well known
\cite{chen2014impossibility} that flat bands of tight-binding models
cannot be energetically isolated and possess nonzero Chern number: coming from a tight-binding model in reciprocal space, the real-space texture in dark OFLs must be degenerate with a different
bright texture at least at one point in the unit cell. This makes the
adiabatic limit more subtle as the gap must be due to the zero-point
motion of the atom. As noted in \cite{nascimbene2024emergence} for
different dark OFLs, the bright-state gap generically vanishes quadratically
around the gapless point. If $V_1$ is the characteristic size of the
potential, the bright state zero-point energy is $\propto \sqrt{M V_1}$. As $V_1\to \infty$ the model will project
perfectly onto the adiabatic subspace of dark states, and be governed
by the adiabatic Hamiltonian \eqref{eq:ham_adiabatic} with geometric
potential contribution $D(\vb r)/2M$. For most dark OFLs this potential
is significant as $D(\vb r)\geq|B_\mr{texture}(\vb r)|$, hence $D(\vb
	r)/2M$ is generically a fluctuating potential comparable to the natural
cyclotron energy of the problem. This means that even as $V_1\to \infty$
a generic dark OFL will \emph{not} lead to a flat Chern band, even
in the absence of diabatic corrections, which decay with the reciprocal
zero-point energy $\propto 1/\sqrt{V_1}$. As our earlier considerations show, while precluding exact flat bands, perturbations of this magnitude to the AC limit largely preserve ideality. This matches the observations that dark state OFLs are very close to ideality, which was noted in \cite{burba2025} and \cite{nascimbene2024emergence}.

\begin{figure}
    \centering
    \includegraphics{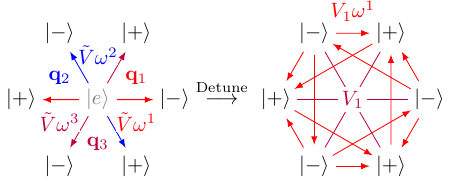}
    \caption{Detuning the excited state by $\delta$  in a three-state dark OFL results in a two-state OFL. Two original couplings $v_1,v_2$ with momenta transfers $\vb q_1$ and $\vb q_2$ result in a new coupling with amplitude  $v_1\bar{v}_2/\delta$ and momentum transfer $\vb q_1-\vb q_2$.}
    \label{fig:detuning}
\end{figure}

We consider a novel dark OFL dual to the tight-binding model
in \cite{yang2025fractional}. This model
corresponds to a dark lattice with uniform flux in reciprocal
space. Let $\Delta,\tilde{V}$ be potential
parameters, and $\ket{e}$ an excited state above the two ground states $|\alpha = \pm\rangle$. In particular  we define the optical potential
\begin{equation}
	\hat{V}(\vb r)=-\Delta\ket{e}\bra{e}+\tilde{V}\sum_{a=1,\alpha=\pm}^3
	\e^{-\i \alpha\vb q_a\cdot \vb
		r-\frac{2\i a\pi}{3}}\ket{\alpha}\bra{e}+\mr{h.c.}\label{eq:pot_darkinit}
\end{equation} 
The (unnormalized) bright state $|s\rangle$, with coefficients $\langle \alpha |s\rangle \equiv s^\alpha(\vb
	r)=\sum_a\e^{-\i\alpha \vb q_a\cdot \vb r-2\pi\i a/3}$ is
orthogonal to the (unnormalized) dark state $|d\rangle$ with $\langle \alpha |d\rangle \equiv d^\alpha(\vb
	r)=\alpha\sum_a\e^{-\i \alpha \vb q_a\cdot \vb r+2\pi\i a/3}$, which
is an exact zero mode of \eqref{eq:pot_darkinit}. When $|\Delta|\gg
	|\tilde{V}|$ the excited state is far detuned and can be eliminated
using perturbation theory, to yield the effective potential ${V}(\vb r)=V_1
	s^\alpha(\vb r)\bar{s}_{\beta}$, where $V_1=\tilde{V}^2/\Delta$. As
illustrated in Fig.  \ref{fig:detuning}.
The precise model we consider adds to this dark OFL a scalar lattice of strength $V_0$, to give a total potential
\begin{equation}
	(V^\Lambda)^\alpha_\beta(\vb r)=V_1s^\alpha \bar{s}_\beta+\delta^\alpha_\beta V_0\sum_{a}\left[\e^{\i \vb G_a\cdot \vb r}+\e^{-\i\vb G\cdot \vb r}\right]\label{eq:pot_dark} \,,
\end{equation}
where $\vb G_2 = \vb q_2-\vb q_1$ etc. as before.
Due to the vanishing potential of the ground state texture, the difference between the adiabatic Hamiltonian \eqref{eq:ham_adiabatic} and the AC Hamiltonian \eqref{eq:ham_ac} is $U(\vb r)=\frac{1}{2M}[D(\vb r)+B_\mr{texture}(\vb r)]$. Remarkably, this is a constant, so this model possesses an exact flat band in the adiabatic limit. To see this choose a coordinate system so that $q_{ax}+\i q_{ay}=|\vb q|\e^{2\pi\i a/3}$, from which it follows that $(\partial_x+\i\partial_y)d^\alpha=-\i|\vb q|\sum_a\e^{-\i \alpha \vb q_a\cdot \vb r+4\pi\i a/3}=-\i|\vb q|s^\alpha$, and since $\braket{-|s}=\braket{d|+}, \braket{+|s}=-\braket{d|-}$ their normalizations agree $|d|^2=|s|^2$. We will also need the orthogonal projector $Q^\alpha_\beta=\delta^\alpha_\beta-\frac{d^\alpha\bar{d}_\beta}{|d|^2}$, which clearly satisfies $Q^\alpha_\beta d^\beta=\bar{d}_\alpha Q^\alpha_\beta=0$ and $s^\alpha=Q^\alpha_\beta s^\beta$. As such
\begin{align}
             \label{eq:qsq1}
	U(\vb r) & =\frac{1}{2M}\left[(\partial_x-\i\partial_y)\frac{\bar{d}_\alpha}{|d|}\right] Q^\alpha_\beta\left[(\partial_x-\i\partial_y)\frac{{d}^\beta}{|d|}\right]  \\
    &=\frac{1}{2M|d|^2}\left[(\partial_x-\i\partial_y)\bar{d}_\alpha\right] Q^\alpha_\beta\left[(\partial_x-\i\partial_y){{d}^\beta}\right]\\
    &=\frac{\vb q^2|s|^2}{2M|d|^2} \\
	         & =\frac{\vb q^2}{2M}.
             \label{eq:qsq4}
\end{align}
Hence in the adiabatic limit $V_1\to\infty$ there must be an exactly flat band with energy $\vb q^2/2M$.
{We emphasize that this is an exact analytical result: the dark-state texture of this OFL satisfies $D(\vb r)+B_\mr{texture}(\vb r)=\mathrm{const}$, which is precisely the condition for the adiabatic Hamiltonian to reduce to AC form.}
In Fig.~\ref{fig:phase-diag} we show the model's phase diagram. Even at $V_0=0,V_1=180\omega_c$ the bandwidth is $5\times 10^{-2}\omega_c$, which is far larger than that of the dual Haldane model. By contrast, adding a tiny scalar potential of magnitude $V_0\sim0.5\omega_c^{3/2}/\sqrt{V_{1}}$ we find a $1$-flat manifold whose bandwidth at $(V_1,V_0)=(181,0.038)\omega_c$ is $1\times 10^{-3}\omega_c$, while the trace violation achieves a minimum value of $2\times 10^{-6}$ at $(V_1,V_0)=(50,0.048)\omega_c$. The strength of the requisite scalar potential follows from the fact that it must cancel diabatic corrections, while the quoted coefficient is found by an empirical least squares fit to the $1$-flat manifold. This model reveals that reaching the flat bands is more subtle than simply canceling the potential deviation $U(\vb r)$ from the AC limit -- the diabatic corrections shift the location of the $1$-flat manifold, even as they do not destroy it. Notably though, a tunable scalar lattice is sufficient to reach the $1$-flat manifold, even though there cannot be any direct cancellation between it and the diabatic perturbation to the Hamiltonian. Furthermore, since the deviation from the AC limit is much smaller in magnitude than the other examples, the band is much closer to ideality.

{Beyond the cold-atom setting, the exact AC equivalence found here suggests a concrete design goal for multilayer TMD heterostructures: rather than only approximating the AC form, as in twisted bilayers, one can ask which tunneling structure would realize it exactly. We note that the dark OFL potential of Eq.~\eqref{eq:pot_darkinit} could be implemented in trilayer transition metal dichalcogenide heterostructures, identifying the state $\ket{e}$ with the central layer and the states $\ket{\alpha=\pm}$ with the two adjacent layers, twisted by $\pm\theta$ relative to the central layer and displaced in-plane in opposite directions by $\Delta x=\pm a_m/\sqrt{3}$, where $a_m$ is the moir\'e lattice constant. The required tunneling structure --- each outer layer coupling only to the central layer, with no direct tunneling between the two outer layers --- is natural for such a trilayer, and would provide a solid-state realization of the same dark-lattice construction, yielding a genuinely (rather than approximately) ideal Chern band. Similar trilayer TMD heterostructures have been explored, for other reasons, in Ref.~\cite{choi2025helical}. We leave a detailed study of this trilayer realization to future work.}

\section{Strongly Correlated Phases}
\label{sec:strongly}

\begin{figure}
    \centering
    \includegraphics[width=1.\linewidth]{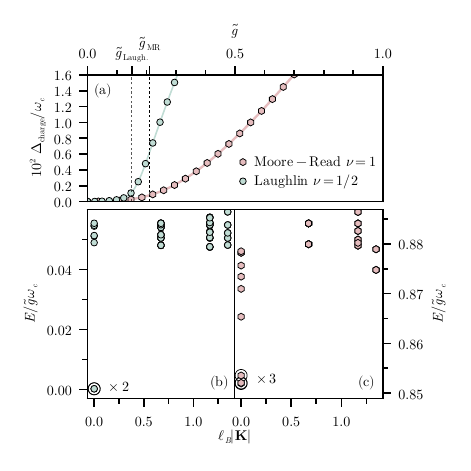}
    \caption{Many-body characteristics of the dual Haldane model at parameters $(V_z,V_+,V_0)=(3.06,7.06,2.09
)\omega_c$ at fillings $\nu=1/2,1$. The charge gap $\Delta_\mr{charge}$ (a) computed with $(N_B,N_\phi)=(8,16),(12,12)$. The exact diagonalization spectra at filling $\nu=1/2$ (b) and $\nu=1$ (c) absent dispersion computed with $(N_B,N_\phi)=(8,16),(16,16)$ respectively. The nearly degenerate states are circled and their multiplicity noted. The many body crystal momentum $\vb K$ is measured from the origin of the Brillouin zone.}
    \label{fig:many-body}
\end{figure}
Our considerations have yielded a zoo of nearly flat bands, with which one can hope to reach the strongly correlated regime. 
We focus on bosonic atoms, for which cold atom platforms offer the greatest differences from solid state systems\cite{cooperbosons}. Also, the flat manifolds are close to perfectly spin polarized at each point in real space, hence contact interactions are Pauli-suppressed for fermionic isotopes (though these remain nonzero\cite{cooperOpticalFluxLattices2011}). Ignoring the species dependence of the scattering length $a_s$, the effective two-dimensional interaction takes the form
\begin{equation}
	V^\mr{int}(\vb r_1-\vb r_2)=\frac{\tilde{g}}{M}\delta^{(2)}(\vb r_1-\vb r_2)
\end{equation}
where a vertical confinement with oscillator length $a_z$ yields $\tilde{g}=\sqrt{8\pi}a_s/a_z$.
We consider $N_B$ bosons occupying a band of $N_\Phi$ states, having filling factor $\nu=N_B/N_\Phi$. The natural energy scale for the interaction in the LLL is the zeroth Haldane pseudopotential strength $v_0={\tilde{g}\omega_c}/{4\pi}$, the size of which sets the typical many body gap for quantum Hall phases. At the coarsest level, the competition between a Bose superfluid and a strongly correlated phase will fall to the latter's favor when $v_0$ exceeds the bandwidth. For a $1$-flat band, given the proximity to ideality, an optimistic estimate is therefore that strong correlations can be reached when $\tilde{g}$ is larger than the characteristic value
\begin{equation}
    \tilde{g}_\mr{char}\sim4\pi|I_{\vb G\in\Lambda^\star_{2}}|{=}\begin{cases}
        0.05&\text{$C_6$ $2\pi$-flux}\\ 
        0.0002&\text{$C_6$ $\phantom{2}\pi$-flux}\\
         0.5&\text{$C_4$ $2\pi$-flux}\\
         0.02&\text{$C_4$ $\phantom{2}\pi$-flux}
    \end{cases}\label{eq:gchar}
\end{equation}
Experimentally, interactions as strong as $\tilde{g}\sim 0.6$ have been realized\cite{Panagiotis2021}, making all options bar the $C_4$ $2\pi$-flux models appealing experimental candidates. To reach $\tilde{g}\sim 0.26$ in $^{87}\mr{Rb}$\cite{Ville2017}, where the scattering length is $a_s\simeq100 a_\mr{Bohr}$, the vertical trap frequency needs to be $\omega=\tilde{g}^2/8\pi Ma_s^2=2\pi\times 10\,\si{\kilo\hertz}$.

For any $N_B$-particle wavefunction in the LLL $\psi^\mr{LLL}(\vb r_1\dots\vb r_{N_B})$ there is a natural ideal analogue
\begin{equation}
    \psi^\mr{ideal}(\vb r_1\alpha_1\cdots)\propto\psi^\mr{LLL}(\vb r_1\cdots)\prod_{i=1}^{N_B}\chi^{\alpha_i}(\vb r_i)\,.
\end{equation}
Ignoring the spinor direction, for model states arising as correlation functions of conformal field theories\cite{MooreR91,HanssonRMPHierarchiesQH2017}, or from the Coulomb gas analogy\cite{laughlinAnomalousQuantumHall1983}, this can be thought of as a modification to the neutralizing background `charge' in those descriptions to be proportional to the effective magnetic field $\lB^{-2}-\nabla^2\phi$. Spinor-independent expectation values $\braket{\cdot}^\mr{ideal}$ from $\psi^\mr{ideal}$ are simply Gibbs reweightings of those $\braket{\cdot}^\mr{LLL}$ from $\psi^\mr{LLL}$\footnote{This has a natural extension to the case of observables which are not diagonal in position basis}
\begin{align}
   \braket{f(\vb r_1\cdots\vb r_{N_B})}^\mr{ideal}=\frac{\Braket{\e^{2\sum_i\phi(\vb r_i)}f(\vb r_1\cdots\vb r_{N_B})}^\mr{LLL}}{\Braket{\e^{2\sum_i\phi(\vb r_i)}}^\mr{LLL}}.
\end{align}
If $f$ is such an observable which is positive definite and for which $\braket{f}^\mr{LLL}=0$ it follows immediately that $\braket{f}^\mr{ideal}=0$ for any non-singular $\phi$.
At filling $\nu=1/2$, the Laughlin wavefunction \cite{laughlinAnomalousQuantumHall1983} of bosons
$\psi^\mr{LLL}_\mr{Laugh.}=\prod_{i<j}(z_i-z_j)^2\e^{-\sum_{i}|z_i|^2/4\lB^2}$ has vanishing interaction energy. Hence, irrespective of $\phi$, the interaction energy will vanish in an ideal band, ensuring that the ideal analogue is a ground state.  
At filling $\nu=1$, the ground state of the bosonic LLL is known to be the Moore-Read Pfaffian\cite{CooperWG00}, possessing model wavefunction $\psi_\mr{MR}^\mr{LLL}=\mr{Pf}(\frac{1}{z_i-z_j})\prod _{i<j}(z_i-z_j)\e^{-\sum_i|z_i|^2/4\lB^2}$. By contrast to the Laughlin state, the interaction energy is non-vanishing due to the finite probability of coincident particles. Using a `mean-field' approximation to describe the change in the correlation functions due to $\phi\neq0$\footnote{In particular, introduce $\phi\to\lambda \phi$. Then $\partial_\lambda\log [\braket{\delta(\vb r-\vb r_1)\delta(\vb r-\vb r_2)\e^{2\lambda\sum_i\phi}}/\braket{\e^{2\lambda\sum_i\phi}}]\simeq 2[2\phi(\vb r)-\partial_\lambda\log\braket{\e^{2\lambda\phi}}]$. The right hand side is a `mean field' approximation to the true change, as it ignores additional correlation induced by the density modulation.}, the ground state energy takes the form
\begin{equation}
    E_\mr{gs}^\mr{ideal}\simeq\frac{2\pi\lB^2\int\dd \vb r\,\e^{4\phi}}{\left[\int\dd \vb r\,\e^{2\phi}\right]^2}E_\mr{gs}^\mr{LLL}=\frac{\braket{\e^{4\phi}}_\mr{uc}}{\braket{\e^{2\phi}}_\mr{uc}^2}E_\mr{gs}^\mr{LLL}\label{eq:int_energy}\,.
\end{equation}
At intermediate fillings $1/2< \nu\lesssim6$, the quantum Hall states in the bosonic LLL compete with other states, notably charge-density-ordered stripe phases\cite{Cooperadvances}. Depending on whether their period is commensurate/incommensurate with $\Lambda_\mr{pot}$, these competing charge-density-ordered states could be favored/disfavored by the lattice-scale modulations. Thus we expect the stability of the Moore-Read state to be sensitive to $\phi$, and must be checked on a case-by-case basis. 

To understand how these considerations manifest empirically, we consider the dual Haldane model at the generic $1$-flat parameter point $(V_z,V_+,V_0)=(3.06,7.06,2.09
)\omega_c$ as in \ref{sec:examples}. We perform exact diagonalization on the lowest band of the model on $C_6$ symmetric super-cells containing $N_\phi=12,16$ unit cells. 
Consider first the strong interaction limit, where the bandwidth is artificially set to $0$.
Consistent with the proximity to ideality we find $2$-fold degenerate ground states at filling $\nu=1/2$ ($N_B=6,8$) with near-vanishing energy per particle $E/N_B=(4.8\pm0.2)\times 10^{-3}\tilde{g}\omega_c$, see Fig. \ref{fig:many-body}. The uncertainty quoted is the energy difference between the two lattice sizes.
At filling $\nu=1$ ($N_B=12,16$) we instead find three nearly degenerate ground states. This is consistent with the Moore-Read state. 
The correlation length is sufficiently large that the ground state degeneracy is not clearly developed in non $C_6$ symmetric super-cell geometries. This is likely exacerbated by the aforementioned competing crystalline stripe phase at intermediate fillings.
The $\nu=1$ Moore-Read state in the Landau level has ground state energy $E^\mr{LLL}_\mr{gs}\simeq4.8\times10^{-2}\tilde{g}\omega_cN_B$\cite{Cooperadvances}, whereas it takes the value $E^\mr{ideal}_\mr{gs}=5.3\times 10^{-2}\tilde{g}\omega_cN_B=1.1E_\mr{gs}^\mr{LLL}$ in the ideal band. From single-particle numerics we find $\braket{\e^{4\phi}}_\mr{uc}/\braket{\e^{2\phi}}^2_\mr{uc}=1.1$, so the ground state energy of the Moore-Read state perfectly matches our ideal estimate \eqref{eq:int_energy}. 

The many-body gaps are comparable to those in the LLL. The neutral gap $\Delta_\mr{neutral}$ is computed as the difference in energy between the first excited state and the lowest of the degenerate ground states, while the charge gap $\Delta_\mr{charge}$ is obtained from the ground state energies with $N_B,N_B\pm1$ particles as:
\begin{equation}
    \hspace{-0.2cm}\frac{\Delta_{\mr{charge}}}{N_B}=\frac{E_\mr{gs}(N_B+1)}{N_B+1}+\frac{E_\mr{gs}(N_B-1)}{N_B-1}-\frac{2E_\mr{gs}(N_B)}{N_B}
\end{equation} 
At half filling $\Delta_\mr{neutral}^{\nu=1/2}\simeq (4.9\pm0.1)\times10^{-2}\tilde{g}\omega_c$ and $\Delta^{\nu=1/2}_\mr{charge}=(7.5\pm0.6)\times 10^{-2}\tilde{g}\omega_c$ which are indistinguishable from those in the LLL with the same particle number. For a typical experimental value of the cyclotron frequency $\omega_c= 2\pi\times 1\,\si{\kilo\hertz}$ and $\tilde{g}=0.6$, the resulting charge gap is therefore $2\pi\times(45\pm4)\,\si{\hertz}$. In Fig. \ref{fig:many-body} we furthermore see that the neutral excitations have lowest energy near $\lB|\vb K|\sim 1$, which is consistent with the typical location of the magneto-roton minimum\cite{magnetoroton}, indicating that dynamics in this phase are very similar to that in the LLL.
At unit filling we find $\Delta^{\nu=1}_{\mr{neutral}}=(1.3\pm0.1)\times10^{-2}\tilde{g}\omega_c$ which is smaller than the neutral LLL gap of $1.8\times10^{-2}\tilde{g}\omega_c$ we compute for $N_\phi=12$. We compute the charge gap $\Delta_\mr{charge}^{\nu=1}=2.5\times 10^{-2}\tilde{g}\omega_c$ at $N_\phi=12$, which is also smaller than in the LLL where it takes the value $4.5\times 10^{-2}\tilde{g}\omega_c$. While smaller than the LLL gaps, they remain large in absolute terms. Furthermore $\Delta_\mr{neutral}$ increases 
between $N_B=12\to16$, giving us confidence that the Moore-Read state survives to the thermodynamic limit. 
To estimate the necessary $\tilde{g}$ to reach the phase transition from the Bose superfluid at $\tilde{g}=0$, we compute the charge gap as a function of $\tilde{g}$, as shown in Fig. \ref{fig:many-body}. In the Bose superfluid the charge gap vanishes, while it increases linearly as a function of $\tilde{g}$ in the strongly correlated phase. Finite size effects smear the transition, though for the Laughlin state at $\nu=1/2$ (computed using $N_B=8$) there is a very clear transition around $\tilde{g}_\mr{Laugh.}\simeq0.15$, where the slope $\dd\Delta_\mr{charge}/\dd \tilde{g}$ reaches half the asymptotic value, which we will use as our criterion to characterize the transition point in other cases. At $\nu=1$ ($N_B=12$) the slope $\dd\Delta^{\nu=1}_\mr{charge}/\dd\tilde{g}$ reaches half its asymptotic value at $\tilde{g}_\mr{MR}\simeq0.2$, but we caution that the transition point of the Moore-Read state is ill-resolved due to the larger correlation length at accessible system sizes.
In Appendix~\ref{app:ED} we show further exact diagonalization results supporting the existence of the Laughlin and more broadly Jain states at $\nu=1/2,2/3,3/4$ in the dual Haldane, triangular, and square models, their charge gap across the superfluid to Laughlin state transition, and further comparisons with the LLL ED spectra. For the dual triangular (square) models at a $1$-flat parameter point with bandwidth $W^\mr{1-flat}_\triangle=3\times10^{-5}\omega_c$ ($W^\mr{1-flat}_\square=10^{-2}\omega_c$) we find a transition with $\tilde{g}^\mr{1-flat}_\triangle=3\times 10^{-4}$ ($\tilde{g}_{\square}^\mr{1-flat}=0.2$), in good agreement with \eqref{eq:gchar}. Since all of these models are essentially ideal, it is reasonable that the transition point is simply proportional to the bandwidth, which is supported by the fact that the aforementioned $2$-flat parameter point in the dual square model has bandwidth $W^\mr{2-flat}_\square=3\times10^{-5}\omega_c$, and transitions to a Laughlin state at $\tilde{g}_{\square}^\mr{2-flat}=5\times 10^{-4}$. {Comparing the $1$- and $2$-flat dual square models, which differ in bandwidth by three orders of magnitude but are otherwise the same model, isolates the role of the bandwidth itself in controlling the onset of the strongly correlated phase.} While the ground state degeneracy is less clearly developed in $\hat{V}^\triangle$ at $\nu=1$, we do find appealing evidence of the Moore-Read state's existence. With $N_B=12$ bosons, there are $3$ putative ground states separated from any excited states by a neutral gap $2\times10^{-2}\tilde{g}\omega_c$, and furthermore, the low-energy spectrum near-perfectly matches the lowest $50$ eigenvalues in the LLL in every momentum sector, differing on average only $1.5\%$ (after subtracting the ground state energy).
\section{Discussion and Conclusion}
We have shown that optical flux lattices can generically support nearly flat and ideal bands through the addition of a suitable scalar potential. By tuning one potential parameter for lattices with $C_6$ $2\pi$-flux and $C_{4,6}$ $\pi$-flux symmetries, one can reach $1$-flat bands, for which the interaction strength $\tilde{g}$ can readily be made big enough in experiment, to reach the Laughlin, Jain sequence, and Moore-Read states. {Although a $1$-flat band already suffices for this purpose, the value of tuning to still narrower bands is mainly conceptual: it gives a controlled starting point from which the departure from the ideal limit can be studied quantitatively via the scalar potential strength, rather than being fixed by material parameters as in moiré systems.} {A particularly sharp result within this framework is the dark-state OFL of Sec.~\ref{sec:darkofl}, whose adiabatic Hamiltonian is \emph{exactly} of Aharonov-Casher form. This is, to our knowledge, the first exact physical realization of the AC structure with nonzero Chern number, in any platform. The AC Hamiltonian has been established as a theoretical ideal~\cite{MoralesPRL2024MATTMD,shiAdiabaticApproximationAharonovCasher2024}, and the present construction provides a concrete, tunable experimental system in which that ideal is achieved precisely.}
Due to the flexibility of this approach, the principal challenge remains to realize a suitable host OFL potential. In alkali-metal gases this is complicated by the fact that the dominant contributions to the polarizability typically arise from the $\mr{D}_{1,2}$ lines, with opposite vector polarizability. To craft a state-dependent lattice potential, the laser frequency must be suitably close to these transitions, which necessarily begets photon scattering. However, for heavy atoms like Rb or Cs, this photon scattering need not be prohibitive for the lifetime. In Appendix \ref{app:DualHaldaneExp} we give examples of experimental setups that can realize the dual Haldane and square models. The photon scattering rate in these setups is of order $1-6\times 10^{-4}\omega_c$. For $^{87}\mr{Rb}$, a wavelength of $\lambda=810\,\si{\nano\meter}$ yields a photon scattering rate around $0.8-1.6\,\si{\second^{-1}}$. In the case of $^{133}\mr{Cs}$, with red detuned beams at wavelength $\lambda=950\,\si{\nano\meter}$, it can be as low as $0.1\,\si{\second^{-1}}$. The photon scattering can be further reduced using other atomic species (including Dy, Er, or Ti) where state-dependent Raman coupling can be achieved with far-detuned light. Instead of using far-detuned light, it is interesting to note that,  absent many-body effects, photon scattering from (near) resonant levels can in principle be reduced arbitrarily in dark lattice setups by increasing the laser power, though to our knowledge, experimentally achieved dark-lattice lifetimes remain small\cite{darkstateexperiment}. To shape higher-shell Fourier components of the lattice, one particularly interesting technique, is the realization of optical potentials using programmable holographic techniques\cite{holographic}. Beyond photon scattering, we caution that magnetic field noise will likely pose an important experimental obstacle, as it contributes a non-uniform perturbation to the system. {Reaching a target filling additionally requires controlling the particle density, which can be tuned via the depth and shape of the confining potential; incompressible states are comparatively easy to target, since the many-body gap locks the density to its quantized value over a finite range of chemical potential, with a Mott-insulator gap in a superlattice offering one likely route to loading a low-entropy state at fixed density before evolving into the topological phase.}
{Once prepared, the topological character of these phases can be probed in several complementary ways, including Hall drift measurements and probes of edge modes and bulk fractionalized quasiparticles. In the near term, the most direct route is likely to be the measurement of multi-particle correlation functions, which characterize the many-body wavefunction itself\cite{readWavefunctionMicroscope2003}. This approach has recently been used to identify a lattice Moore-Read state, via a suppression of short-range three-body density correlations together with a Hall drift measurement\cite{kwan2026pfaffian}.}
Irrespective of atomic species, we have provided a general approach by which to realize flat ideal Chern bands and robust fractional quantum Hall states. The ability to precisely tune band geometry will enable systematic studies of the transition into exotic topologically ordered phases, including the non-Abelian states sought for fault-tolerant quantum computation. Furthermore, this approach strengthens the connection between cold atoms and moiré materials, promising a new, highly controllable platform to probe the fundamental principles of strongly correlated systems and test theoretical models of ideal band structures in a clean, controllable, and, well-characterized experimental environment. {While the two-state OFLs studied here, built from the Aharonov-Casher form, are restricted to bands of unit Chern number, reciprocal-space lattice  are known to realize bands of higher Chern number\cite{cooper2012designing}; we anticipate that a similar scalar-potential tuning could flatten such bands, though we have not explored this here.}

{\it Note added:} After completion of this work we learned that the authors of Ref.~\cite{nascimbene2024emergence} have now found a square version of a dark OFL which, using an analysis mirroring (\ref{eq:qsq1}-\ref{eq:qsq4}), one can show shares the property of (\ref{eq:pot_darkinit}) that it has an exact flat band in the adiabatic limit. For finite lattice depth, it is similarly amenable to band-flattening using scalar potentials on $\Lambda^*_{1,2,\ldots}$ momentum shells.

\section{Acknowledgments}
We thank Patrick Ledwith, Tomohiro Soejima, Monika Aidelsburger, Dan Parker, Ashvin Vishwanath, Alexander Douglas, and Clemens Kuhlenkamp for helpful discussions, and Jean Dalibard for helpful discussions and for sharing an updated version of Ref.~\cite{nascimbene2024emergence}.
This work was supported by a Simons Investigator Award (Grant No. 511029) and the Engineering and Physical Sciences Research Council [grant numbers EP/V062654/1 and EP/Y01510X/1]. O.E.S. was funded by NSF DMR-2220703.

\newpage
\appendix
\begin{figure*}[tp]
    \centering
    \includegraphics{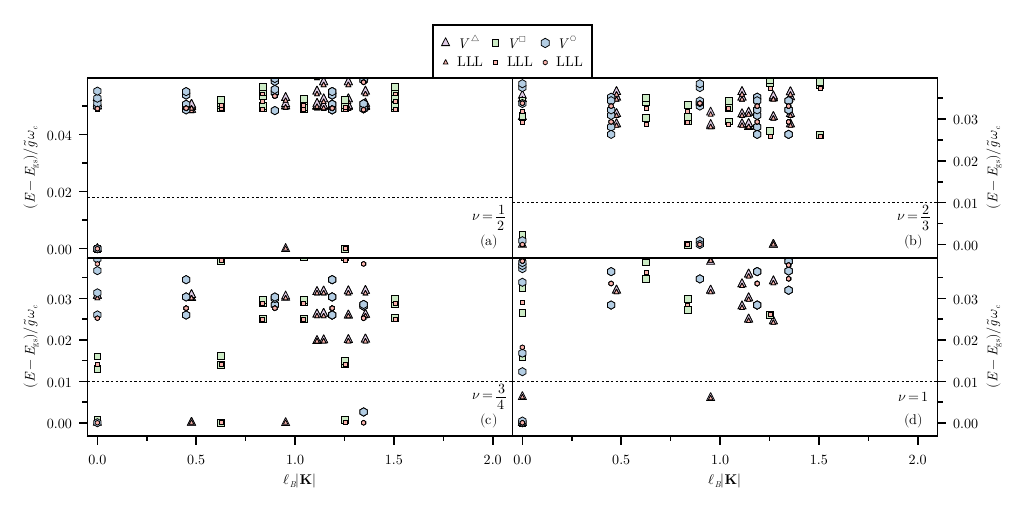}
    \caption{Many body ED spectra of dual Haldane, square and triangular models in the dispersionless limit, along with that for the LLL in the same lattice geometries. The dashed lines separate ground states from excited states, and we note that we have subtracted the ground state energy in all cases. The panels correspond to (a) $\nu=1/2,N_B=6$, showing $2$-fold ground state degeneracy (b) $\nu=2/3,N_B=8$ showing $3$-fold ground state degeneracy (c) $\nu=3/4, N_B=9$ showing $4$-fold ground state degeneracy and (d) $\nu=1,N_B=12$ showing $3$-fold ground state degeneracy.}
    \label{fig:ED_extra}
\end{figure*}

\section{Further Exact Diagonalization Results}
\label{app:ED}

Beyond what is shown in the main text, we emphasize that the existence of FQH states is generic to our models, and their spectra closely resemble those of the LLL.
Consider Figure~\ref{fig:ED_extra}, where dispersionless exact diagonalization results for $N_\phi=12$ and $N_B=6,8,9,12$ are shown, in the first 3 cases for each of the $1$-flat dual Haldane, triangular and square models, and in the last case only for the dual Haldane and triangular models. Alongside these we show the LLL spectrum in the same super-cell geometries, and as can be clearly observed, the agreement is remarkable. This is especially so in the case of the dual square and triangular models, while there are visible splittings in the case of the dual Haldane model. The calculations are carried out for $V^\triangle$ at parameters $(V_0,V_1)=(1.39,10.55)\omega_c$, and for $V^\square$ at $(V_+,V_z,V_0)=(4.88,4.04,0)\omega_c$, while $V^{\hexagon}$ is at the aforementioned parameter point with a $C_6$ symmetry super-cell. In all models we see clear many-body gaps for $\nu=1/2,2/3,3/4$ with ground state degeneracy $2,3,4$ consistent with the bosonic Jain sequence at filling $\nu=p/(p+1)$ for $p=1,2,3$, which is further supported by the near-exact match between the model and LLL spectra. At filling $\nu=1$ the emergent composite fermions feel no average effective magnetic field. There is a competition between charge ordering into stripes, and pairing, the latter of which results in the Moore-Read state. This competition  and the large correlation length are evident from how sensitive the spectra are to the super-cell geometry. For the dual Haldane model, in a $C_6$ symmetric super-cell, a clear 3-fold degenerate ground state occurs at the Brillouin zone origin. By contrast, the triangular model does display 3 low-lying states, but they are far from degenerate. In the square case no ground state manifold is apparent in any super-cell geometry we have tried, so we do not display the spectra. Still, in all cases, especially the square and triangular models, the spectra are a near-exact match for those of the LLL, for which it is known that the ground state is the Moore-Read state. Thus we have confidence that the split degeneracy is a finite size effect, and that the physics of the LLL survives essentially unmodified in these ideal bands.

\begin{figure}
    \centering
    \includegraphics{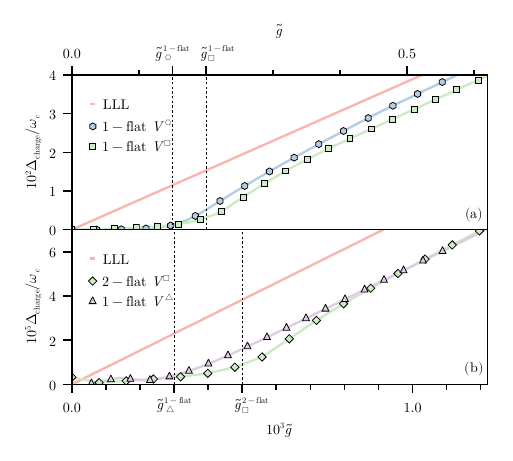}
    \caption{Charge gap for the superfluid-Laughlin state transition at $\nu=1/2$ computed for $N_B=8$ bosons. We also include the genuine LLL charge gap, which is directly proportional to $\tilde{g}$.}
    \label{fig:chargegap_extra}
\end{figure}

The phase transitions between the superfluid and FQH states are governed by the competition between the single particle and interaction energies. We study $4\times 4$ supercells with $N_B=8$ bosons in the lowest band in the aforementioned $1$-flat models, along with the $2$-flat dual square model at parameters $(V_+,V_z,V_0)=(4.19,3.77,0.87)\omega_c$. As can be seen in Fig. \ref{fig:chargegap_extra}, the phase transition to the Laughlin state occurs in the $1$-flat dual Haldane and square models around $\tilde{g}\simeq0.15-0.2$ while it occurs in the $2$-flat dual square and the $1$-flat dual triangular model at the much smaller $\tilde{g}\simeq(3- 5)\times 10^{-4}$. While this is interesting from the point of view of reaching strong correlations, it is likely that effects beyond the bandwidth would limit the observation of the Laughlin state for too small many body gaps, which puts an effective lower limit on $\tilde{g}$ regardless of the transition point.
\section{Experimental schemes}
\label{app:DualHaldaneExp}
For state-dependent optical lattices, the choice of two internal states has large experimental implications on the preparation and lifetime of the experiment. In keeping with the nature of the flat bands, our considerations here are general, and should be considered merely an example for a possible experimental procedure. Consider a ground state manifold characterized by definite combined orbital and spin angular momentum $J$ and nuclear spin $I$, so that in the absence of magnetic field the manifold splits further into levels of definite total angular momentum $F=|I-J|,\dots,I+J$. In the case of alkali-metal atoms, the ground state has $J=1/2$. For fermionic isotopes $F$ is half-integer while for bosonic isotopes it is integer. Furthermore, applying a small quantizing magnetic field, which we take to be in the vertical $z$ direction, the energy levels may be labeled $\ket{F,m_F}$.

Increasing the magnetic field until the quadratic Zeeman effect detunes the atomic transitions by energies in large excess of $\omega_c$, a two-level system can be furnished from a pair $\ket{\pm}=\ket{F_{\pm},m_{\pm}}$. Let us focus on the case of off-resonant lasers, so that the optical lattice potential arises from the second order light shift. This is controlled by the dynamic polarizability tensor operator $\alpha_{ab}$ of the ground state manifold, that is characterized by a combination of scalar $\alpha_S$, vector $\alpha_V$, and tensor $\alpha_T$ components. For each frequency $\omega$, this decomposition becomes
\begin{align}
	\alpha_{ab}(\omega) & =\alpha_S(\omega)\delta_{ab}-\alpha_V(\omega)\frac{\i\epsilon_{abc}\hat{J}_c}{2J}                                            \nonumber \\
	                    & \phantom{=}+\alpha_T(\omega)\frac{3[\hat{J}_a\hat{J}_b+\hat{J}_b\hat{J}_a]-2\delta_{ab}\hat{\vb J}^2}{2J(2J-1)}\label{eq:pol}
\end{align}
where for notational clarity the $\hat{}$ have been added on angular momentum operators. Generically $\alpha_{ab}(\omega)$ is predominantly real, but has a finite imaginary component. This is the remnant of the finite life time of the excited states which the light field mixes with the ground state. 
\subsection{Alkali Metals}
Projected to within the $\ket{\pm}$ subspace, the polarizability need not take a form as elegant as \eqref{eq:pol}. Let us initially focus on the situation where $\alpha_T=0$, as is appropriate for alkali-metals, notably Rubidium and Cesium.
Since the vector polarizability allow for off-diagonal matrix elements differing by at most $1$ unit of angular momentum, the states $\ket{\pm}$ must be chosen such that $|m_+-m_-|\leq 1$ to have any hope for creating an OFL. Naturally this leads to two qualitatively different kinds of OFL designs coherent with the choice of schemes
\begin{enumerate}[label={V\arabic*}]
	\item \label{scheme:S1} $m_+=m_-+1$, while $F_{\pm}=F$ ($F_+\neq F_-$ is perfectly possible but we don't consider it).
	\item \label{scheme:S2}$m_+=m_-=m$ while $F_\pm=I\pm1/2$
\end{enumerate}
Scheme \ref{scheme:S1} has the advantage of not suffering two-body hyperfine relaxation in collisions, but we often find simpler setups using scheme \ref{scheme:S2}, and in the Laughlin state, collisions are strongly suppressed, likely making hyperfine relaxation an unimportant source of atomic loss.
In the first scheme \ref{scheme:S1}, all three angular momentum operators have non-trivial projector onto the two-level system
\begin{align}
	\hat{J}_3     & \overset{\text{\ref{scheme:S1}}}{\mapsto} 2J( A+B\hat{\sigma}_3) \\
	\hat{J}_{1,2} &  \overset{\text{\ref{scheme:S1}}}{\mapsto}2J C \hat{\sigma}_{1,2}
\end{align}
for three constants $A,B,C$. Ignoring the mixing caused by finite magnetic field, these coefficients can be easily worked out 
\begin{align}
B&=\frac{F(F+1)+J(J+1)-I(I+1)}{8JF(F+1)}\\
    A&=(2m_-+1)B\\ 
    C&=B\sqrt{F(F+1)-m_-m_+}
\end{align}
Importantly for bosonic isotopes, there is no possible choice of levels within scheme \ref{scheme:S1}, such that $A=0$ for small magnetic fields. This conclusion is not altered by considering states from different $F_+\neq F_-$. For fermionic isotopes such a pair of states could be $m_F=\pm1/2$.
This form leads to the projected polarizability
\begin{equation}
	\alpha_{ab}\overset{\text{\ref{scheme:S1}}}{\mapsto}\alpha_S\delta_{ab}-\i A\alpha_V\epsilon_{ab}-\i B\alpha_V\hat{\sigma}_3\epsilon_{ab}-\i C\alpha_V\epsilon_{abc}\hat{\sigma}_{c\perp}
\end{equation}
which shows the unfortunate reality that the vector shift $\hat{\sigma}_3$ will be accompanied by a scalar component. This scalar component is furthermore phase-shifted relative to the usual scalar component, which often leads to undesirable effects, and can make it hard to cancel.
Consider for example the laser setup of Ref.~\cite{cooperOpticalFluxLattices2011} realizing the dual Haldane model when $A=0$. The two levels are split in energy by $\delta$, and the setup consists of three planar lasers with frequency $\omega$, amplitude $\mc{E}$ and polarization angle $\theta$ to the vertical, as well as a vertical circularly polarized laser of frequency $\omega+\delta$ and amplitude $r\mc{E}$
\begin{align}
	\vb E_\omega          & =\mc{E}\sum_a(\cos\theta\vb e_z+\sin\theta \hat{\vb q}_a\wedge\vb e_z)\e^{\i\vb q_a\cdot \vb r} \\
	\vb E_{\omega+\delta} & =r\mc{E}\e^{\i|\vb q|\vb e_z\cdot \vb r}\frac{\vb{e}_x+\i \vb e_y}{\sqrt{2}}
\end{align}
In the rotating frame, the total potential takes a form $V^{\hexagon}+V^\mr{const.}$ with parameters:
\begin{align}
	{V}_0    & =-\frac{\alpha_S(3\cos^2\theta-1)-\alpha_V\i\sqrt{3}A\sin^2\theta}{2}\mc{E}^2                                                                    \\
	{V}_+    & =-\alpha_VCr\sqrt{2}\cos\theta\mc{E}^2                                                                                                           \\
	{V}_z    & =\frac{\alpha_VB\sqrt{3} \sin^2\theta}{2}\mc{E}^2\\
    V^\mr{const.}&=-(3+r^2) \alpha_S\mc{E}^2  
\end{align}
Recall the illustration of the dual Haldane model in Fig. \ref{fig:reciprocal_spaceillustration}. When $V_0$ is not purely real, the two sublattices in the dual Haldane model experience unequal reciprocal space flux. This breaks the $C_6$ symmetry, making it impossible to reach the $1$-flat manifold by tuning a single parameter. When $A\neq 0$ there is no way to avoid this with this particular setup.
Under scheme \ref{scheme:S2}, we find that 
\begin{align}
    J_3\overset{\text{\ref{scheme:S2}}}{\mapsto }\tilde{C}\hat{\sigma}_1+\tilde{B}\hat{\sigma}_3\qquad J_{1,2}\overset{\text{\ref{scheme:S2}}}\mapsto0
\end{align} 
where 
\begin{align}
    |\tilde{C}|&=\frac{\sqrt{(I+1/2)^2-m^2}}{2I+1}\\
    \tilde{B}&=\frac{m}{2I+1}
\end{align}
While $J_{1,2}\mapsto 0$ does impose some design limitations, they are not so severe as to prevent the creation of OFL.
Under \ref{scheme:S2} the polarizability takes the form 
\begin{equation}
    \alpha_{ab}\overset{\text{\ref{scheme:S2}}}{\mapsto}\alpha_S\delta_{ab}-\i \tilde{B}\hat{\sigma}_3\alpha_V\epsilon_{ab}-\i \tilde{C}\alpha_V\hat{\sigma}_1\epsilon_{ab}
\end{equation}
Independent control of the different vector components of the potential are possible since only resonant beams contribute to the optical potential, but more complicated optical setups than previously considered are often still necessary.
\subsection{Dual Haldane Model for bosons in scheme \ref{scheme:S1}}
One possibility to realize the dual Haldane model is by using multiple frequencies for the planar lasers. Such a multi frequency scheme was practically realized using electro-optical modulators in \cite{kosch2022multifrequency}. In particular, consider a scheme involving a main frequency $\omega$, along with sidebands $\omega_{1,2,3}$ and an out-of-plane laser with frequency $\omega+\delta$, as illustrated in Fig. \ref{fig:dual_haldane_setup}
\begin{figure}
	\begin{tikzpicture}
		\node (alabel) at (-2.25,1.5) {(a)};
		\node (blabel) at (1.8,1.5) {(b)};
		\node (a) at (-0.75,0){
			\tdplotsetmaincoords{75}{130} % rotate 60 degrees around x axis, then 105 degrees about z
			\begin{tikzpicture}[tdplot_main_coords,scale=0.8]

				\begin{scope}[canvas is xy plane at z=0,rotate=00]
					\draw[black,opacity=0.15] (0,0) circle (2.4);
					\iffalse
						\foreach\i in {1,2,...,12}{
								\draw ({1.5*cos(30*\i-15)},{1.5*sin(30*\i-15)})--({1.5*cos(30*\i+15)},{1.5*sin(30*\i+15)});
							}
					\fi
					\foreach\i in {1,2,3}{
							\draw[thick,-latex,red] ({2.3*cos(120*\i-120)},{2.3*sin(120*\i-120)})-- node[xshift=-8, above=0.28] {$ \omega,\vb q_{\i}$} ({0.4*cos(120*\i-120)},{0.4*sin(120*\i-120)});
						}

				\end{scope}

				\draw[black,-latex,thick] (0,0,-1.25) -- node[below right]{$\omega+\delta$} (0,0,-0.4);
				\draw[yscale=0.4,thick,decoration={						markings,
							mark=at position 0.45 with {\arrow{latex}}},postaction=decorate](0,0,-2) +(205:0.35) arc(205:-125:0.35);

				\foreach\i in {0,8}{
						\draw[-latex,blue,dashed] ({2*cos(30*\i)},{2*sin(30*\i)},0.1)-- ({0.5*cos(30*\i)},{0.5*sin(30*\i)},0.1);
					}
				\foreach\i in {4,8}{
						\draw[-latex,purple,dashed] ({2*cos(30*\i)},{2*sin(30*\i)},0.2)-- ({0.5*cos(30*\i)},{0.5*sin(30*\i)},0.2);
					}
				\foreach\i in {0,4}{
						\draw[-latex,green,dashed] ({2*cos(30*\i)},{2*sin(30*\i)},0.3)-- ({0.5*cos(30*\i)},{0.5*sin(30*\i)},0.3);
					}

			\end{tikzpicture}
		};
		\node (c) at (3.5,0){
			\begin{tikzpicture}[scale=0.8]
				% Draw all levels
				\draw[level] (0,0) -- node[above] {$\ket{+}$} (1,0);
				\draw[level] (1.25,-0.5) -- node[above] {$\ket{-}$} (2.25,-0.5);
				\draw[level] (0,1) -- node[above ] {$\mathrm{P}_{1/2}$}(2.25,1);
				\draw[dashed,level,red] (0,2) -- node[above] {}(1,2);
				\draw[dashed,level,red] (1.25,1.5) -- node[above] {}(2.25,1.5);

				\draw[dashed,level,thin,blue] (0,2.1) -- node[above] {}(1,2.1);
				\draw[dashed,level,thin,blue] (1.25,1.6) -- node[above] {}(2.25,1.6);

				\draw[dashed,level,thin,purple] (0,2.2) -- node[above] {}(1,2.2);
				\draw[dashed,level,thin,purple] (1.25,1.7) -- node[above] {}(2.25,1.7);

				\draw[dashed,level,thin,green] (0,2.3) -- node[above] {}(1,2.3);
				\draw[dashed,level,thin,green] (1.25,1.8) -- node[above]
				{}(2.25,1.8);)
				\draw[level] (0,3) -- node[below] {$\mathrm{P}_{3/2}$}(2.25,3);
				\draw[dashed,|-|] (2.45,-0.5)--node[right]{$\omega$}(2.45,1.5);
				\draw[dashed,|-|] (-0.2,-0.5)--node[left]{$\delta$}(-0.2,0);

				% Draw labels
			\end{tikzpicture}};
	\end{tikzpicture}
	\caption{(a) The imagined
		optical setup for the dual Haldane model in bosonic atoms under scheme \ref{scheme:S1}. There is one circularly polarized out-of-plane laser, and
		three main in-plane lasers. Each main laser has two different side bands as
		indicated by the color, which are detuned by an electro-optical modulator as in \cite{kosch2022multifrequency}. See equation \eqref{eq:S1full} (b) Level diagram showing
		the two internal states, the virtual levels dressed by the laser frequencies, as well as the excited $P$ levels from which the polarizability derives.
	}
	\label{fig:dual_haldane_setup}
\end{figure}
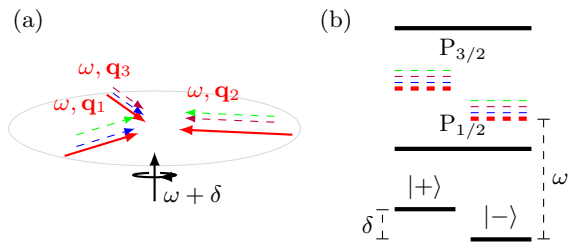
\begin{align}
	\vb E_\omega          & =\mc{E}\sum_a(\cos\theta\vb e_z+\sin\theta \hat{\vb q}_a\wedge\vb e_z)\e^{\i\vb q_a\cdot \vb r} \\
	\vb E_{\omega_a}      & =r_\parallel\mc{E} \vb e_z [\e^{\i \vb q_{a+1}\cdot \vb r+\i\phi}+\e^{\i\vb q_a\cdot \vb r}]    \\
	\vb E_{\omega+\delta} & =r\mc{E}\e^{\i|\vb q|\vb e_z\cdot \vb r}\frac{\vb{e}_x+\i \vb e_y}{\sqrt{2}}.\label{eq:S1full}
\end{align}
There is great flexibility in how to distribute the scalar potential between the side and main frequency, with simplest being that $\phi=\pi/2$ (or $-\pi/2$ corresponding to the sign of $\alpha_V A/\alpha_S$), and that
\begin{equation}
	\alpha_S r_\parallel^2=\frac{\sqrt{3}\sin^2\theta A \alpha_V}{2}
\end{equation}
Then the parameters $V_0,V_+,V_z$ of the dual Haldane model become
\begin{align}
	V_0   & =-\frac{3\cos^2\theta-1}{2}\alpha_S \mc{E}^2      \\
	V_+ & =r\sqrt{2}\cos\theta C\alpha_V \mc{E}^2           \\
	V_z   & =\frac{\sqrt{3}\sin^2\theta}{2} B\alpha_V\mc{E}^2
\end{align}
The overall constant of the potential has an imaginary part given by
\begin{equation}
	\mr{Im} V^\mr{const.}=
	\left[-(3+r^2)-3^{3/3}\sin^2\theta A \frac{\alpha_V}{|\alpha_S|} \right]\mr{Im}{\alpha}_S\mc{E}^2
\end{equation}
This term is the dominant source of heating in the lattice when $\alpha_S\gg \alpha_V$. In this regime, the scalar lattice strength is largely irrelevant for the heating calculation, and the beams must be close to orthogonal so that $\cos\theta\sim1/3$. The overall heating is dominated by the uniform background potential
\begin{equation}
	\mr{Im} V^\mr{const.}\approx-\left(3\sqrt{3}+\frac{B^2 V_+^2}{C^2 V_z^2}\right)\frac{V_z}{|B|}\frac{\mr{Im}{\alpha_S}}{\alpha_V}
\end{equation}
For any particular parameter point of the dual Haldane model, the far detuned heating is limited by the ratio $\Im\alpha_S/\alpha_V$ which tends to $3\Gamma/4(\omega_{3/2}-\omega_{1/2})$, see Appendix \ref{app:pol} for details. 
Consider the $1$-flat parameter point from \ref{sec:examples}, where $(V_z,V_+,V_0)=(3.06,7.06,2.09
)\omega_c$. The heating is similar for any choice of $m_\pm$, so for simplicity let $m_-=-F$. Then the inverse lifetime for 
\begin{align}\Gamma^\mr{\gamma}&\simeq\left(3\sqrt{3}+\frac{5.3}{2I-1}\right)\frac{9.2\omega_c\Gamma (2I+1) }{\omega_{3/2}-\omega_{1/2}}\\
&=\frac{\Gamma\omega_c}{\omega_{3/2}-\omega_{1/2}}\times\begin{cases}
        144&I=3/2\\
        180&I=5/2\\ 
        223&I=7/2
    \end{cases}
\end{align}
For $^{87}\mr{Rb}$ with $I=3/2$ and using the values from \ref{tab:Dline}, we find $\Gamma^\mr{asymp}=5.6\times10^{-4}\omega_c$. At the red-detuned wavelength $\lambda=810\,\si{\nano\meter}$ we have $\omega_c=2\pi\times2.9\,\si{\kilo\hertz}$, so $\Gamma^\mr{asymp}\simeq1.6\,\si{\second^{-1}}$, while for $^{133}\mr{Cs}$ where $I=7/2$ a red detuned laser of wavelength $\lambda=950\,\si{\nano\meter}$ we have $\omega_c=2\pi\times 1.3\,\si{\kilo\hertz}$, while $\Gamma^\mr{asymp}=3.1\times10^{-4}\omega_c$, and hence $\Gamma^\mr{asymp}\simeq0.4\,\si{\second^{-1}}$. We note that while the heating rate for Rubidium is significantly larger, which is primarily due to the increased recoil energy. As such, for a fixed $\tilde{g}$ the ratio between the many body gaps and the decay rate, which is what ultimately determines the stability of the state, differs much less in the two cases.

\subsection{Dual Haldane Model in scheme \ref{scheme:S2}}
Alternatively, we can pick a manifold of constant $m_F$ but containing two different $F_\pm=I\pm1/2$, so that $J_3\mapsto \tilde{C}\hat{\sigma}_1+\tilde{B}\hat{\sigma}_3,J_{1,2}\mapsto0$. In this case, consider a laser setup with two main frequencies $\omega,\tilde{\omega}$ for in-plane lasers, and a single out-of-plane laser at frequency $\omega+\delta$:
\begin{align}
	\vb E_\omega
	 & =\mc{E} \sum_a\e^{\frac{2\pi\i a}{3}}
	(\cos\theta\vb{e}_z
	+\sin\theta \hat{\vb q}_a\wedge\vb e_z)
	\e^{\i\vb q_a\cdot \vb r}                   \\
	\vb E_{\tilde{\omega}}
	 & =\mc{E}\sum_a\e^{-\frac{2\pi \i a}{3}}
	(\cos\theta\vb{e}_z
	+\sin\theta \hat{\vb q}_a\wedge\vb e_z)
	\e^{\i\vb q_a\cdot \vb r}                   \\
	\vb E_{\omega+\delta}
	 & =r\mc{E}\e^{\i|\vb q|\vb e_z\cdot \vb r}
	\frac{\vb{e}_x+\i \vb{e}_y}{\sqrt{2}}.
\end{align}
This results in the lattice potential of a dual Haldane model with 
\begin{align}
	V^\mr{const} & =-(6+r^2) \alpha_S\mc{E}^2                                                                                       \\
	                        V_0&=-(3\cos^2\theta-1)\alpha_S\mc{E}^2                           \\
	                        V_+& {=}-r\sqrt{2} \cos\theta \tilde{C}\alpha_V\mc{E}^2\\
	                        V_z&= \frac{\sqrt{3}\sin^2\theta \tilde{B}\alpha_V}{2}\mc{E}^2
\end{align}
Again, it is the constant piece of the potential that is the biggest source of heating when $\alpha_S\gg\alpha_V$,
\begin{align}
  \Im  V^\mr{const.} & =-\left(6+\frac{V_+^2\tilde{B}^2}{2V_z^2\tilde{C}^2}\right)\frac{V_z\sqrt{3}\Im\alpha_S}{\tilde{B}\alpha_V}
\end{align}
For $m_-=-F$ we find for the same parameters as before in $^{87}\mr{Rb}$ a heating rate $\Gamma^\mr{asymp}=2\,\si{\second^{-1}}$, while for $^{133}\mr{Cs}$ we find $\Gamma^\mr{asymp}=0.2\,\si{\second^{-1}}$.

\subsection{Dual Square model under scheme \ref{scheme:S2}}
Consider the optical lattice generated by a light field with frequencies $\omega,\omega+\delta$ where $\delta$ is the splitting between $\ket{\pm}$. We consider a quartet of wavevectors $\vb q_a=q\vb e_x\cos \tfrac{\pi a}{2} +q\vb e_y\sin\tfrac{\pi a}{2}$ where the $a$ index is an integer mod $4$.
The electric field components take the form
\begin{align}
    \vb E_\omega&=\frac{\mc{E}}{\sqrt{2}}\sum_a \e^{\frac{\i \pi a}{2}}(\e^{\i\phi}\vb e_z+\hat{\vb q}_{a+1})\e^{\i\vb q_a\cdot \vb r}\\ 
    \vb E_{\omega+\delta}&=\frac{r\mc{E}}{\sqrt{2}}\e^{\i|\vb q|\vb e_z\cdot \vb r} (\vb e_x+\vb e_y)
\end{align} 
where $\phi$ is a free parameter that does not impact the resulting potential, but may make experimental design more straightforward.
The resulting scalar potential is a constant
\begin{align}
    V^\mr{const.} &=(4+|r|^2)\mc{E}^2\alpha_S
\end{align}
while the spatially dependent potential takes the form of the $\pi$-square model \eqref{eq:pot_square} with 
\begin{align}
V_z&=\mc{E}^2\alpha_V\tilde{B} \\ 
V_+&=\frac{r\mc{E}^2\alpha_V\tilde{C}}{2}\\ 
V_0&=0
\end{align}
Even without utilizing the additional scalar parameter we can reach the $1$-flat condition. An additional scalar potential as prescribed by the model can be included using beams of wavevector $2\vb q_a$ at frequency $2\omega$. When the electric field has the form 
\begin{equation}
    \vb E_{2\omega}=\frac{\tilde{r}{\mc{E}}}{\sqrt{2}}\sum_a(\e^{\i\tilde{\phi}}\vb e_z+\hat{\vb q}_{a+1})\e^{2\i\vb q_a\cdot \vb r}
\end{equation}
for some common phase $\e^{\i\tilde{\phi}}$. This additional light field creates a scalar potential of the form 
\begin{equation}
   \alpha_S(2\omega)\tilde{r}^2\mc{E}^2\left(4+\sum_a\e^{2\i(\vb q_{a+1}-\vb q_a)}\right)
\end{equation}
while the interference cancels any vector shift. The resulting model is then the $\pi$-square model with 
\begin{align}
V^\mr{const.}&=\mc{E}^2([4+r^2]\alpha_S(\omega)+4\tilde{r}^2\alpha_S(2\omega))\\
    V_z&=\mc{E}^2\tilde{B}\alpha_V(\omega) \\ 
V_+&=\frac{r\mc{E}^2\tilde{C}\alpha_V(\omega)}{2}\\ 
V_0&=\mc{E}^2\tilde{r}^2\alpha_S(2\omega)
\end{align}
In the regime of interest $\alpha_S\gg \alpha_V$, while $V_0\sim V_+$, so $\tilde{r}\ll 1$, and the additional heating from the scalar lattice is completely negligible. The heating rate takes the form
\begin{equation}
    \Gamma^\gamma\simeq-\frac{8V_z\Im\alpha_S}{\tilde{B}\alpha_V}\left[1+\frac{V_+^2\tilde{B}^2}{V_z^2\tilde{C}^2}\right]
\end{equation}
At the $2$-flat parameter point $(V_+,V_z,V_0)=(4.19,3.77,0.87)\omega_c$, we find an asymptotic heating rate of $\Gamma^\mr{asymp}\simeq0.8\,\si{\second^{-1}}$ at $\lambda=810\,\si{\nano\meter}$ in $^{87}\mr{Rb}$, while at $\lambda=950\,\si{\nano\meter}$ in $^{133}\mr{Cs}$ it becomes $\Gamma^\mr{asymp}\simeq0.1\,\si{\second^{-1}}$.

\section{Landau level review}
\label{app:landau}
For convenience we give a short review of Landau levels on a plane in the periodic basis\cite{Haldane1985}.
Consider the usual affine plane $\R^2$ with coordinates $x^1,x^2$ and magnetic field
$F_{ab}$, which is related to the flux
density $B$ by the anti-symmetric tensor $F_{ab}=\epsilon_{ab}B$.
We consider identical charge $+1$ particles moving on the plane. When these are transported around the plane, they acquire an Aharonov-Bohm phase. The generator of this translation is the kinetic momentum
$\bm{\pi}$, which has the form $\pi_a=-\i\nabla_a-A_a$ where
$A_a$ is the connection. The operator commutation relations take the form
\begin{equation}
	[x^a,x^b]=0\qquad [x^a,\pi_b]=\i\delta^a_b\qquad [\pi_a,\pi_b]=\i F_{ab}.\label{eq:magalg}
\end{equation}
Even without specifying a particular Hamiltonian, it is natural to ponder the existence of operators that commute with both $\pi_{1,2}$. When $B$ is constant \eqref{eq:magalg} is a central extension of the usual Heisenberg algebra, and we may construct such operators, which are the guiding center momenta
$\tilde{\pi}_a=\pi_a-F_{ab}x^b$. Their name is chosen since $R^a=\epsilon^{ab}\tilde{\pi}_b/B$ are the typical guiding center operators. The two kinds of momenta commute
\begin{equation}
	[\tilde{\pi}_a,\pi_b]=\i F_{ab}-F_{ac}[x^c,\pi_b]=\i F_{ab}-\i F_{ab}=0
\end{equation}
while the guiding center momenta have opposite chirality to $\pi_a$
\begin{equation}
	[\tilde{\pi}_a,\tilde{\pi}_b]=-\i F_{ab}
\end{equation}
The guiding center momenta are distinct from the kinetic momenta, and
as such there are two natural notions of translation operators. The parallel
transport translations
$t(\vb d)=\e^{\i \vb d\cdot \bm\pi}$ and the guiding center translations $\hat{t}(\vb
	d)=\e^{\i\vb d\cdot \hat{\bm \pi}}$. Since they have the same commutation
relations with the position operator $\vb x$, if $V(\vb x)$ is an operator
$t^\dagger(\vb d)V(\vb x)t(\vb d)=\hat{t}^\dagger(\vb
	d)V(\vb x)\hat{t}(\vb d)=V(\vb x-\vb d)$. Their translation algebras are inherited from their
respective momentum algebras
\begin{equation}
	t(\vb d_1)t(\vb d_2)=t(\vb d_1+\vb
	d_2)\e^{-\frac{\i}{2}F_{ab}d_1^ad_2^b}
\end{equation}
whereas
\begin{equation}
	\hat{t}(\vb d_1)\hat{t}(\vb d_2)=\hat{t}(\vb d_1+\vb
	d_2)\e^{+\frac{\i}{2}F_{ab}d_1^ad_2^b}.
\end{equation}
Let us suppose magnetic field or potential inhomogeneities conspire so that only
translations along some lattice $\Lambda=\vb a_1\Z\oplus \vb a_2\Z$ are
symmetries, with a unit cell that encloses $2\pi$ flux $F_{ab}a_1^a a_2^b=2\pi$.
The dual lattice we denote $\Lambda^\star=\vb g_1\Z\oplus \vb g_2\Z$,
where $ \vb g^i_{a}=\epsilon^{ij}F_{ab}\vb a^b_j$.
This ensures that the corresponding guiding center translations commute $\hat{t}(\vb a_1)\hat{t}(\vb a_2)=\hat{t}(\vb a_2)\hat{t}(\vb
	a_1)$, hence they can be simultaneously diagonalized.
This motivates defining their common eigenstates, which are essentially the magnetic Bloch wave-functions, with crystal momenta $\vb k$ by
\begin{equation}
	\hat{t}(\vb a_i)\ket{\vb k}=-\e^{\i \vb a_i\cdot\vb k}\ket{\vb k}\label{eq:transsym}
\end{equation}
in analogy with the usual translation symmetry. The choice of overall $-$ sign
corresponds to a shift in the origin of the Brillouin zone, which is made since
it makes the resulting wave-functions more symmetric between real and momentum
space. 
With this preamble, let us understand the magnetic Bloch wave-functions.
To write
these down explicitly, we must pick a gauge connection. For example the symmetric
gauge wherein
$A_a=-\frac{1}{2}F_{ab}x^b$, and hence
$\tilde{\pi}_a=-\i\nabla_a-\tfrac{1}{2}F_{ab}x^b$. 
Consider first understanding
the eigenstates of $\vb a_1\cdot \hat{\bm{\pi}}$ with eigenvalue $2\pi\phi_1$. These eigenstates
satisfy
\begin{equation}
	2\pi \phi_1 f_{\phi_1}(\vb x)=\left(-\i \vb a_1\cdot \nabla-\frac{\epsilon_{1j}}{2}\vb g^j\cdot\vb
	x\right)f_{\phi_1}(\vb x)
\end{equation}
which can be directly integrated to yield
\begin{equation}
	f_{\phi_1}(\vb x)=\e^{\i \phi_1\vb x\cdot \vb g^1+\frac{\i\epsilon_{12}}{4\pi}(\vb g^2\cdot\vb
		x)(\vb g^1\cdot \vb x)}
\end{equation}
up to a multiplicative function dependent only on $\vb g^2\cdot\vb x$.
Since the guiding center translations by $\vb a_{1,2}$ commute, we can form a
superposition of $f_{\phi_1+n}$ for $n\in \Z$ states that are
$\hat{t}(\vb a_2)$ eigenstates. It is straightforward to compute the action of
$\hat{t}(\vb a_2)$ on $f_{\phi_1}$
\begin{equation}
	\hat{t}(\vb a_2)f_{\phi_1}(\vb x)=\e^{\i \vb g^1\cdot\vb x }f_{\phi_1}(\vb
	x)=f_{\phi_1+1}(\vb x)
\end{equation}
The most general eigenstate of $\hat{t}(\vb a_{1,2})$ with eigenvalues $\e^{\i 2\pi \phi_{1,2}}$ must be a superposition of the form
\begin{equation}
	\chi(\vb x)=\sum_{n\in\Z}f_{\phi_1+n}(\vb x)h_n(\vb g^2\cdot\vb x)
\end{equation}
where $h_n$ is some set of functions.
To find conditions relating $h_n$ we investigate the action of $\hat{t}(\vb a_2)$. In particular
\begin{equation}
	\e^{\i2\pi\phi_2}\chi=   \hat{t}(\vb
	a_2)\chi_{\phi_1,\phi_2}=\sum_nf_{\phi_1+n+1}(\vb x)h_n(\vb g^2\cdot \vb x+2\pi)
\end{equation}
From the linear independence of the summand, it is clear $\e^{\i 2\pi
		\phi_2}h_{n}(\vb g^2\cdot\vb x)=h_{n-1}(\vb g^2\cdot \vb x+2\pi)$, which immediately implies
\begin{equation}
	h_n(\vb g^2\cdot \vb x)=\e^{-\i 2\pi\phi_2 n}\tilde{h}(\vb g^2\cdot \vb x+2\pi n)
\end{equation}
for an arbitrary function $\tilde h$. In particular let us factor out $\tilde{h}=\e^{\i \phi_2
		\vb g^2\cdot \vb x}h$. In terms of this arbitrary function, the simultaneous eigenfunctions take the form
\begin{equation*}
	\chi_{\phi_1\phi_2}(\vb x)=\e^{\i \phi_i\vb g^i\cdot \vb x+\frac{\i}{4\pi}(\vb g^1\cdot\vb x)(\vb g^2\cdot\vb x)}\sum_{n\in\Z}\e^{\i n\vb g^1\cdot\vb x}h(\vb g^2\cdot \vb x+2\pi n)
\end{equation*}
This is naively very asymmetric between the two lattice vectors, but they can be
related by Poisson resummation, which exchange $\vb g^{1,2}$ and replaces $h$
with its Fourier transform. In particular the Fourier series of the Dirac comb
can be expressed
\begin{align}
	\sum_{n\in \Z}\delta(x-2\pi n) & =\frac{1}{2\pi}\sum_{n\in\Z} \e^{\i n x}                                 \\
	                               & =\sum_{n\in\Z}\int_\R \frac{\dd y}{2\pi} \e^{\i xy/2\pi}\delta(y-2\pi n)
\end{align}
Defining $\hat{h}(y)=\int \dd x\e^{\i xy/2\pi}h(x)/2\pi$ Poisson resummation relates
\begin{align}
	 & \sum_{n\in\Z}\e^{\i n\vb g_1\cdot\vb x}h(\vb g^2\cdot \vb x+2\pi n)                                                      \\
	 & =\int\dd
	\theta\sum_{n\in\Z}\e^{\i \theta\vb g^1\cdot \vb
	x/2\pi}h(\theta)\delta(\theta-2\pi n)                                                                                       \\
	 & =\sum_{n\in \Z}\int \frac{\dd \theta\dd\phi }{2\pi}\e^{\i \theta(\phi+\vb
	g^1\cdot \vb x)/2\pi}h(\vb g^2\cdot \vb x+\theta)\delta(\phi-2\pi n)                                                        \\
	 & =\e^{-\frac{\i}{2\pi}(\vb g^1\cdot \vb x)(\vb g^2\cdot \vb x)}\sum_{n\in \Z}\hat{h}(\vb g^1\cdot \vb x +2\pi n)\e^{-\i n
		\vb g^2\cdot \vb x}
\end{align}
so we could equally well write
\begin{equation}
	\chi_{\phi_1\phi_2}=\e^{\i\phi_i\vb g^i\cdot \vb x-\frac{\i}{4\pi}(\vb
		g^1\cdot \vb x)(\vb x\cdot\vb g^2)}\sum_{n\in \Z}\e^{-\i n\vb g^2\cdot \vb
		x}\hat{h}(\vb g^1\cdot \vb x+2\pi n)
\end{equation}
The ordering of the lattice basis is arbitrary, so the
most natural basis to expand $h$ in would be a self-dual basis under Fourier
transformation. Such a basis is provided by the eigenstates of the  harmonic
oscillator with a free parameter $\tau$ in the complex upper
half plane. They take the form of the Hermite functions
$h_{\tau,N}(x)=H_N\left(\sqrt{\frac{\tau}{2\pi \i}}x\right)\e^{\i\frac{\tau}{4\pi} x^2}$, where $H_N$ is the $N$th Hermite polynomial.
Under Fourier transform we find
\begin{equation}
	\hat{h}_{\tau,N}(y)= \frac{\i^N}{2\pi}H_N\left(\sqrt{\frac{\i}{2\pi
			\tau}}y\right)\e^{-\i\frac{1}{4\pi\tau}
		y^2}=\frac{\i^N}{2\pi}h_{-1/\tau,N}(y)
\end{equation}
The corresponding wavefunctions are
closely related to the Jacobi theta function $\theta(u,\tau)=\sum_{n\in\Z} \e^{\i \pi n^2\tau+\i nu}$ (with periods $2\pi,2\pi\tau$).
In particular for $h_{\tau,0}$:
\begin{align}
	 & \sum_{n\in \Z}\e^{\i n\vb g^1\cdot \vb x}\e^{\i\frac{\tau}{4\pi}(\vb g^2\cdot \vb
	x+2\pi n)^2}                                                                                                 \\
	 & =\sum_{n\in\Z}\e^{\i\pi\tau n^2 +\i n\vb x\cdot ( \vb
		g^1+\tau \vb g^2)+\i\frac{\tau}{4\pi}(\vb g^2\cdot
	\vb x)^2}                                                                                                    \\
	 & =\theta\left(\vb x\cdot(\vb g^1+\tau \vb g^2);\tau\right)\e^{\i\frac{\tau}{4\pi}(\vb g^2\cdot \vb x)^2} 
\end{align}
Since the complex numbers come with a natural metric, a choice for the free parameter $\tau$ amounts to picking a metric on the plane. We complexify coordinates using the map:
\begin{equation}
	z=\frac{\lB}{\sqrt{2\pi\Im\tau}}\vb x\cdot(\vb g^1+\tau\vb
	g^2)
\end{equation}
Under this map $a_1=\lB\sqrt{\frac{2\pi}{\Im\tau}}$ and $a_2=\lB\tau\sqrt{\frac{2\pi}{\Im\tau}}$, so the area of the unit cell is 
\begin{equation}
    \frac{1}{2}|a_1\bar{a}_2-\bar{a}_1a_2|=\frac{\lB^2\pi}{\Im\tau}|\tau-\bar{\tau}|=2\pi\lB^2
\end{equation}
To convert from $\phi$ to $\vb k$ in the wavefunction, recall the minus sign of equation \eqref{eq:transsym}. First consider $\phi_1=\phi_2=0$ which corresponds to a corner of Brillouin zone:
\begin{align}
  \Theta(z;\tau) &= \e^{\frac{\i}{4\pi}[(\vb g^1+\tau\vb g^2)\cdot
			\vb x]\vb g^2\cdot \vb x}\theta\left(\vb x\cdot(\vb g^1+\tau \vb g^2);\tau\right)\\&=\e^{\i  z\Im
	z/2\lB^2}\theta\left(\frac{\sqrt{2\pi \Im\tau} }{\lB} z;\tau\right)                
\end{align}
To center the wavefunction in the Brillouin zone translate it by $\vb a_1/2+\vb a_2/2$ to obtain 
\begin{align}
    \braket{\vb x|\vb k=\vb 0}&\propto\e^{\frac{\i}{4}{(\vb g^1-\vb g^2)\cdot \vb x}}\Theta(z+a_1/2+a_2/2)\\ 
    &\propto\sigma(z)\e^{-|z|^2/4\lB^2}
\end{align}
where $\sigma(z)$ is the holomorphic (modified) Weierstrass sigma function\cite{haldane2018modular} of $\Lambda$.
The $\vb k\neq \vb 0$ case can be constructed by similar translations:
\begin{align}
	\braket{\vb x|\vb k} & =\mc{N}\e^{-\frac{1}{4\lB^2}|z- \i \lB^2k|^2+\frac{\i}{2}\vb
	k\cdot \vb x}\sigma(z-\i\lB^2k)                                              \\
	                     & =\mc{N}\e^{-\frac{1}{4}(\lB^{-2}|z|^2+\lB^2|k|^2-2\i \bar{k}z)}\sigma(z-\i\lB^2k)
\end{align}
where $\mc{N}$ is a $\vb k$ independent normalization constant whose value is not important. In fact $\braket{\vb x|\vb k}$ is the lowest Landau level wavefunction on a torus.
\subsection{Higher Landau levels and form factors}
\label{app:lllform_factor}
Another way of understanding the construction is to consider an operator that
commutes with $\hat{t}(\vb a_{1,2})$, and
split the $\vb k$ sector by the spectrum of this operator. A natural choice
is the kinetic energy $\bm\pi^2$. The $n$th level of $\bm{\pi}^2$ is the $n$th Landau level. Let $z=x+\i y$ and define the corresponding Wirtinger derivatives $\partial z=1;\bar{\partial}z=0$. In terms of these, the complexified kinetic and guiding center momenta take the form
\begin{equation}
	\pi =-2\i\partial+\frac{\i \bar{z}}{2\lB^2};\qquad\tilde{\pi}=\pi- \frac{\i\bar z}{\lB^2}.
\end{equation}
They satisfy the commutation relations 
\begin{equation}
    [\pi,\pi^\dagger]=-2\lB^{-2}\qquad[\tilde{\pi},\tilde{\pi}^\dagger]=2\lB^{-2}
\end{equation}
where all other commutators vanish. After a $\sqrt{2}\lB^{-1}$ rescaling they therefore form two independent harmonic oscillator algebras. Furthermore $\bm \pi^2=\pi \pi^\dagger+2\lB^{-2}$, so we are in effect studying the harmonic oscillator. Note that $\pi^\dagger$ annihilates holomorphic functions multiplying the Gaussian 
$\e^{-\lB^{-2}|z|^2/4}$ of which the wavefunction $\braket{\vb r|\vb k}$ is an example. The general Hermite functions from before are precisely the states that can be found by applying $\pi$ to $\ket{\vb k}=\ket{n=0,\vb k}$, namely
\begin{equation}
	\ket{n,\vb k}=\frac{\lB^n\pi^n}{\sqrt{2^nn!}}\ket{\vb k}
\end{equation}
We use the guiding center translations to change momentum sectors:
\begin{equation}
    \ket{n,\vb k}=\e^{\lB^{2}(\bar{k}\tilde{\pi}^\dagger-{k}\tilde{\pi})/2}\ket{n,\vb 0}=\hat{t}(\vb k_\perp)\ket{n,\vb 0}
\end{equation}
hence the action of the translation operators becomes
\begin{align}
	\hat{t}(\vb k_1^\perp)\ket{n,\vb k_2}
	                                      & =\e^{\frac{\lB^2}{4}(\bar{k}_1k_2-\bar{k}_2k_1)}\ket{n,\vb k_1+\vb k_2}
\end{align}
Noting the Bloch-condition, we deduce boundary conditions for the wavefunctions
\begin{align}
	\ket{n,\vb k+\vb g^{i}} & =\e^{\frac{\lB^2}{4}(\bar{k}g^i-\bar{g}^ik)}\hat{t}((\vb g^i)^\perp)\ket{n,\vb
	k}                                                                                   \\
	                      &= -\e^{\frac{\lB^2}{4}({k}\bar{g}^i-{g}^i\bar{k})}\ket{n,\vb
	k}   
\end{align}
This basis is very suitable for numerical calculations.
The position operator takes the form $z=\i\lB^2(\pi^\dagger-\tilde{\pi}^\dagger)$ and $\bar{z}=\i\lB^{2}(\tilde{\pi}-\pi)$.
This implies that the plane wave operator factorizes
\begin{equation}
	\e^{\i \vb q\cdot  \vb x}=\e^{\lB^{2}(q\pi-\bar{q}\pi^\dagger)/2}\e^{\lB^2(\bar{q}\tilde{\pi}^\dagger-q\tilde{\pi})/2}
\end{equation}
Denote the canonical displacement operator 
\begin{align}D^n_m(\alpha)&=\braket{n|\e^{\alpha a^\dagger-\bar\alpha a}|m}\\ 
    &=\e^{-|\alpha|^2/2}\begin{cases}
        \sqrt{\frac{n!}{m!}}\alpha^{m-n}L_n^{(m-n)}(|\alpha|^2)&m\geq n\\ 
       \sqrt{\frac{m!}{n!}}\bar{\alpha}^{ n-m}L_m^{(n-m)}(|\alpha|^2)&n\geq m
    \end{cases}\nonumber
\end{align}
where $L^{(a)}_b(x)$ are the associated Laguerre polynomials. With this in hand, suppose $\vb q=\vb k_1-\vb k_2+\vb G$ for some reciprocal lattice vector $\vb G\in\Lambda^\star$. Then the form factor takes the easily computable form:
\begin{align}
	 & \braket{n,\vb k_1|\e^{\i \vb q\cdot \vb x}|m,\vb k_2}                          \\
    &=D^n_m(q\lB/\sqrt{2})\e^{\i\lB^2\vb q\wedge \vb k_2/2}\braket{0,\vb k_1|0,\vb k_1+\vb G}\\
     &=D^n_m(q\lB/\sqrt{2})\eta_{\vb G}\e^{\i\lB^2\vb q\wedge \vb k_2/2+\i\lB^2\vb G\wedge\vb k_1/2}
\end{align}
where $\eta_{\vb G}$ is the parity of the reciprocal lattice vector such that $\eta_\vb G=1$ if $\vb G\in2\Lambda^\star$ and $-1$ otherwise.
We use these matrix elements for numerical computation involving $\tilde{H}$. To do this note that $\psi^\alpha_{n\vb k}(\vb r)$ obeys the Bloch periodicity $\psi^\alpha_{n\vb k}(\vb r+\vb R)=\e^{\i(\vb k+\vb q_\alpha)\cdot\vb R }\psi^\alpha_{n\vb k}(\vb r)$. Factoring out the LLL wavefunction with crystal momentum $\vb k$ means that $\bar\chi$ transforms as  
${\chi}^\alpha(\vb r+\vb R)=-\e^{\i\vb q_\alpha\cdot \vb r-\i\lB^{-2}\vb R\wedge\vb r/2}\chi^\alpha(\vb r)$ -- it obeys the magnetic translation symmetry with opposite chirality to $\braket{\vb r|\vb k}$. Note the invariance of the algebra under the relabeling $\pi\to\tilde{\pi}^\dagger$ and $\pi^\dagger\to \tilde{\pi}$. From this we can reuse the earlier matrix elements, by expanding $\chi^\alpha(\vb r)=\sum_{\bar{n}}c_{n\alpha}\braket{\vb r|\overline{n\vb q_\alpha}}$, where  $\ket{\overline{n\vb k}}$ are the Landau levels of opposite chirality. In particular we define them by the properties $t(\vb a_i)\ket{\overline{n\vb q_\alpha}}=-\e^{\i\vb a_i\cdot \vb  q_\alpha}\ket{\overline{n\vb q_\alpha}}$, $\ket{\overline{n\vb q_\alpha}}=\lB^{n}(\tilde{\pi}^\dagger)^n\ket{\overline{\vb q_\alpha}}/\sqrt{2^nn!}$, and $\braket{\vb r|\overline{\vb k}}=\overline{\braket{\vb r|-\vb k}}$. Including a spinor label we have basis states $\ket{\overline{n\vb k}\alpha}$.
The matrix elements 
\begin{equation}
    \frac{1}{2M}\braket{\overline{n\vb q_\alpha}|\tilde{\pi}\tilde{\pi}^\dagger|\overline{m\vb q_\alpha}}=(n+1)\omega_c\delta^n_{m}\,,
\end{equation}
while for a potential of the form $V^\alpha_\beta(\vb r)=\sum_{\vb G}V^\alpha_\beta(\vb G)\e^{\i(\vb G+\vb q_\alpha-\vb q_\beta)\cdot \vb r}$ (for some $\vb q_\alpha,\vb q_\beta$ which displace the Brillouin zones of the internal states) we have
\begin{align}
  &\hspace{-0.75cm} \braket{\overline{n\vb q_\alpha}\alpha|\hat{V}|\overline{m\vb q_\beta}\beta}\\
   &=\sum_{\vb G}V^\alpha_\beta(\vb G)\braket{\overline{n\vb q_\alpha}|\e^{\i(\vb G+\vb q_{\alpha}-\vb q_\beta)\cdot \vb r}|\overline{m\vb q_\beta}}\nonumber\\
   &=\sum_{\vb G}V^\alpha_\beta(\vb G)\eta_{\vb G}D^{n\alpha}_{m\beta}\left(\vb G\right)\e^{-\i\lB^2\left[\vb G+\frac{\vb q_\alpha-\vb q_\beta}{2}\right]\wedge\left[\frac{\vb q_\alpha+\vb q_\beta}{2}\right]}\nonumber\\
   D^{n\alpha}_{m\beta}\left(\vb G\right)&=D^{n}_m\left([{\bar{G}+\bar{q}_\alpha-\bar{q}_\beta}]\lB/\sqrt2\right)\nonumber\,.
\end{align}
For comparison between the optimal ideal and true wavefunctions we find it most convenient to find the latter by solving \eqref{eq:exact}. To do this we need the matrix elements of $\hat{\zeta}_{\vb k}$ which are
\begin{align}
     &\braket{\overline{n\vb q_{\alpha}}|\hat{\zeta}_{\vb k}|\overline{m\vb q_\alpha}}\\&=-\omega_c\sum_{\vb G\neq \vb 0}G^{-1}\e^{\i\lB^2\vb G\wedge\vb k} \braket{\overline{n\vb q_\alpha}|\e^{\i\vb G\cdot \vb r}\tilde{\pi}^\dagger|\overline{m\vb q_\alpha}}\nonumber\\ 
     &=-\omega_c\sqrt{2(m+1)}\sum_{\vb G\neq\vb 0}\frac{\e^{\i\lB^2\vb G\wedge(\vb k-\vb q_\alpha)}}{\lB G}\eta_{\vb G}D^{n}_{m+1}(\bar{G}\lB/2)\nonumber\,.
\end{align}
Using these pieces we can reproduce the same spectrum from $\tilde{H}+\hat{\zeta}_{\vb k}$ as from a plane-wave expansion of the original Hamiltonian. In practice we truncate the single-particle space to include the first $100$ Landau levels.
\subsection{Trace violation for the $n$th Landau level}
\label{app:landau_trace}
We claimed in the main text that the trace violation for the $n$th Landau level is $2n$, which can be succinctly shown using the technology we have developed so far in this Appendix. The Bloch wavefunctions of the lowest Landau level take the form 
\begin{align}
    u_{0\vb k}(\vb r)&=\e^{-\i\vb k\cdot \vb r}\braket{\vb r|\vb k}\\
    &=
    N\e^{-\frac{1}{4}(\lB^{-2}|z|^2+\lB^2|k|^2)-\i {k}\bar{z}/2}\sigma(z-\i\lB^2k)\,.
\end{align}
Except for the Gaussian $\e^{-\lB^2|k|^2/4}$, this is holomorphic in $k$. The $n$th Landau level Bloch wavefunction can be similarly computed by noting that $\e^{-\i\vb k\cdot \vb r}\pi=(\pi+\bar{k})\e^{-\i\vb k\cdot\vb r}$, hence 
\begin{equation}
    \ket{u_{n\vb k}}=\frac{(\pi+\bar{k})^n}{2^{n/2}\sqrt{n!}}\e^{-\i\vb k\cdot \vb r}\ket{\vb k}\,,
\end{equation}
from which it is immediately clear that
\begin{equation} \partial_{\bar{k}}\ket{u_{n\vb k}}=\lB\sqrt{\frac{n}{2}}\ket{u_{n-1\vb k}}-\frac{\lB^2k}{4}\ket{u_{n\vb k}}\,.
\end{equation}
The Berry curvature is uniform for each Landau level:
\begin{align}
    \Omega(\vb k)&=2\partial_{k}\braket{u_{n\vb k}|\partial_{\bar{k}}u_{n\vb k}}-2\partial_{\bar{k}}\braket{u_{n\vb k}|\partial_{{k}}u_{n\vb k}}\\
    &=4\partial_{k}\braket{u_{n\vb k}|\partial_{\bar{k}}u_{n\vb k}}\\
    &=-\lB^2\,.
\end{align}
We see that it is always negative and results in Chern number $\mc{C}=-1$.
The corresponding trace violation is straightforwardly computed as
\begin{align}
    \tr g+\Omega&=4\bra{\partial_ku_{n\vb k}}(1-\ket{u_{n\vb k}}\bra{u_{n\vb k}})\ket{\partial_{\bar{k} }u_{n\vb k}}\\
    &=2n\lB^{2}\braket{u_{n-1\vb k}|u_{n-1\vb k}}\\ 
    &=2n\lB^{2}\,.
\end{align}
Hence $(2\pi)^{-1}\int\dd \vb k (\tr g+\Omega)=2n$ as claimed.
\subsection{Circular dichroism criteria}
\label{app:circular_dichroism}
A natural alternative to the trace condition as a measure of proximity to ideality relates to the null vectors of the quantum geometric tensor $q^{ab}(\vb k)=g^{ab}(\vb k)+\i\epsilon^{ab}\Omega(\vb k)/2$. For ideal bands (with sufficient rotational symmetry) either $w_a=(\vb e_x+\i \vb e_y)_a/\sqrt{2}$ or its complex conjugate $\bar{w}_a$ are null-vectors of $q^{ab}$ for all $\vb k$. Consider measuring the magnitude
\begin{align}
   || q^{ab}w_b||^2&=\left(g^{ab}\bar{w}_b-\frac{\i\epsilon^{ab}\bar{w}_b\Omega}{2}\right)\left(g^{ac}w_c+\frac{\i\epsilon^{ac}w_c\Omega}{2}\right)\nonumber\\
   &=\frac{1}{4}(\Omega-\tr g)^2+\frac{1}{4}(\tr g)^2-\det g\,.
\end{align}
Using $\frac{1}{4}(\tr g)^2\geq\det g$ and taking $\Omega<0$ we find 
\begin{align}
    \frac{||q\bar{w}||^2}{||qw||^2}&=\frac{(\tr g+\Omega)^2+[(\tr g)^2-4\det g]}{(\tr g-\Omega)^2+[(\tr g)^2-4\det g]}\\
    &\geq\left(\frac{\tr g+\Omega}{\tr g-\Omega}\right)^2\,,
\end{align}
or in short
\begin{equation}
    {\frac{\sqrt{2\tr g(\tr g+\Omega)}}{\tr g- \Omega}}\geq\frac{||q\bar{w}||}{||qw||}\geq \frac{\tr g+\Omega}{\tr g-\Omega}\,.
\end{equation}
Ignoring the covariance of the denominator and the numerator, and to leading order in the trace violation, the average of this ratio can be approximately bounded 
\begin{equation}
   \sqrt\frac{T}{2|\mc{C}|} \gtrsim\int\frac{\dd \vb k}{2\pi\lB^{-2}}\frac{||q\bar{w}||}{||qw||}\gtrsim\frac{T}{2|\mc{C}|}\,.
\end{equation}
Whenever the trace violation is small, so too must this average ratio be, but precise numerical comparison between the two quantities is not possible.  This makes direct comparison between the deviation from ideality cited in \cite{nascimbene2024emergence} with the trace violations we quote practically difficult, though qualitatively they show similar features. For example, it monotonically increases in the $n$th Landau level
\begin{equation}
  \frac{||q\bar{w}||}{||qw||}={\frac{n}{n+1}}
\end{equation}
though the rate is much gentler than the trace condition.
\section{Fourier expansion of $\hat{\zeta}$} 
In the main text we claimed the following identity regarding the periodic completion of the Weierstrass $\zeta$ function\cite{haldane2018modular}
\begin{equation}
    \hat{\zeta}(z)=\zeta(z)-\bar{z}/2\lB^2=-\i\sum_{\vb G\neq\vb 0}\frac{\e^{\i\vb G\cdot\vb r}}{\lB^2G}\,.
\end{equation}
The most straightforward derivation is to consider the anti-holomorphic derivative of this function:
\begin{equation}
    \bar\partial({\zeta}-\bar{z}/2\lB^2)=\pi\sum_{\vb R\in\Lambda}\delta^2(\vb r)-1/2\lB^{2}
\end{equation}
where the Dirac comb follow from the pole locations of $\zeta$. 
The Dirac comb can be Fourier expanded, showing that 
\begin{equation}
    \bar{\partial}\hat{\zeta}=\frac{1}{2\lB^2}\sum_{\vb G\neq \vb 0}\e^{\i\vb G\cdot \vb r}\,.
\end{equation}
Periodicity implies that there is a Fourier expansion with coefficients $c_\vb G$ of $\hat{\zeta}$
\begin{equation}
    \hat{\zeta}(z)=\sum_{\vb G}c_{\vb G}\e^{\i\vb G\cdot\vb r}
\end{equation}
hence taking derivatives of this expression and matching terms we find $\i G c_{\vb G}/2=1/2\lB ^2$ for $\vb G\neq \vb 0$. Since $\hat{\zeta}(z)$ has vanishing average, the claimed Fourier expansion follows.
\section{Polarizability of Alkali Atoms}
\label{app:pol}
For completeness, let us derive the polarizability of alkali-metal atoms in the ground state. We label the internal atomic states, including the excited states by $\alpha$. In the dipole approximation, an external electric field of strength $
	\vb E(t)=\sum_\omega \vb E_\omega\e^{-\i\omega t}$ modifies the Hamiltonian by
\begin{equation}
	\delta H=-\vb d\cdot \vb E(t)=\sum V_{\beta}^\alpha(\omega)\e^{-\i\omega t}\ket{\alpha}\bra{\beta}
\end{equation}
where $V^{\alpha}_\beta(\omega)=-\vb d^{\alpha}_\beta\cdot \vb E_\omega$.
If the internal levels are governed by a reference Hamiltonian $H_0=\sum \omega_\alpha\ket{\alpha}\bra{\alpha}$, the perturbation in the $H_0$ rotating frame becomes
\begin{equation}
	\delta H^\mr{rot}=\sum_{\alpha\beta\omega} V^\alpha_\beta(\omega)\e^{-\i(\omega-\omega_\alpha+\omega_\beta)t}\ket{\alpha}\bra{\beta}\,.
\end{equation}
To recast this as a static problem, we use a time-dependent Schrieffer-Wolff transformation $\ket{\psi}=\e^{-S(t)}\ket{\tilde{\psi}}$ where
\begin{equation}S(t)=-\sum_{\alpha\beta\omega}\frac{V^\alpha_\beta(\omega)\e^{-\i (\omega-\omega_\alpha+\omega_\beta)t}}{\omega-\omega_\alpha+\omega_\beta}\ket{\alpha}\bra{\beta}.
\end{equation}
Let $H_\mr{atom}=\sum E_\alpha\ket{\alpha}\bra{\alpha}$ be the true atomic Hamiltonian absent driving, and pick $H_0$ so that $|E_\alpha-\omega_\alpha|\ll |\omega_\alpha-\omega_{\beta}|$ for all $\alpha\neq\beta$ that are coupled.
To second order in $\vb E$, the time evolution of $\ket{\tilde{\psi}}$ is governed by 
\begin{equation}
    H^{(2)}=\sum_\alpha (E_\alpha-\omega_\alpha)\ket{\alpha}\bra{\alpha}+\delta H^{(2)}
\end{equation}
where corrections to from the Schrieffer-Wolff transformation to the atomic piece have been neglected due to the proximity between $H_\mr{atom}$ and $H_0$, and where  
\begin{align}
	\delta H^{(2)} & =-\frac{1}{2}\sum_{\alpha\beta\gamma\omega_1\omega_2}\ket{\alpha}\bra{\beta}\e^{-\i(\omega_1+\omega_2-\omega_\alpha+\omega_\beta)t}               \nonumber                                                                            \\
	               & \phantom{=}\times\left[\frac{V^\alpha_\gamma(\omega_1)V^\gamma_\beta(\omega_2)}{\omega_1-\omega_\alpha+\omega_\gamma}  -\frac{V^\alpha_\gamma(\omega_1)V^\gamma_\beta(\omega_2)}{\omega_2-\omega_\gamma+\omega_\beta} \right]\,.
\end{align}
Whenever the resonance condition $\omega_1+\omega_2-\omega_\alpha+\omega_\beta=0$ is met, this results in a static term. We assume the rotating wave approximation where only these resonant terms are important, that is ignoring higher order light shifts and taking any off-resonant oscillations much faster than the dynamics of interest. Due to the reality of $\vb E(t)$, clearly $\vb E_{-\omega}=\bar{\vb E}_{\omega}$. Furthermore, we consider the situation where the energy splitting of the states of interest is much smaller than the frequencies $\omega$, so that the only resonant contributions arise when $\omega_1<0<\omega_2$ or conversely, and by focusing on ground states furthermore neglecting the contribution to the polarizability from virtual levels formed below the ground state. Combining these considerations, the energy takes the form
\begin{align}
	\delta H^{(2)}
	 & =-\sum_{\omega_1,\omega_2>0,\mr{res}}\ket{\alpha}\bra{\beta}\frac{(\bar{\vb E}_{\omega_1}\cdot \vb d^\alpha_\gamma) (\vb E_{\omega_2}\cdot\vb d^\gamma_\beta)}{\omega_\gamma-\omega_2-\omega_\beta} \\
	 & =-\sum_{K,q}(-)^{K-q}\alpha^{K,-q}[\bar{E}(\omega_1)\otimes E(\omega_2)]^{K,q}
\end{align}
where in the last line, we have transformed to the spherical tensor basis with $K=0,1,2$ corresponding to scalar, vector, and tensor polarizabilities. This picture is particularly clear when the lifetimes of all states are infinite so that $H_\mr{atom}$ is Hermitian. Unfortunately this is not true for the excited states where the imaginary part $\Im E_\alpha=-\Gamma_\alpha/2$ measures the line-width. To account for this $H_0$ should include the linewidth for states which are not resonant, so that $\omega_\alpha\mapsto \omega_\alpha-\i\Gamma_\alpha/2$ is complex. The mismatch between the true and reference energies $E_\alpha-\omega_\alpha$ functions as a species-dependent chemical potential due to the lasers being off-resonant. 
\subsection{Spherical tensor decomposition}
The electrons in the atom have orbital angular momentum $\vb L$, under which $\vb d$ is a vector operator, as well as spin angular momentum $\vb S$. When the atomic reference Hamiltonian $H_0$ is rotationally symmetric, eigenstates carry definite angular momentum $\vb J=\vb L+\vb S$. Consider a degenerate manifold of excited states $\gamma$ with definite angular momentum $J_\gamma$, coupling to ground states with angular momentum $J$. We can work out the $K=0,1,2$ polarizabilities using the Wigner-Eckart theorem, which is to say using the $\mr{SU}(2)$ representation theory.  We use the convention that the reduced matrix elements of a spherical tensor is related to the $3j$ symbol as follows: 
\[
	\braket{J_1m_1|T^{Kq}|J_2m_2}=(-)^{J_1-m_1}\braket{J_1||T^K||J_2}\begin{pmatrix}
		J_1 & K & J_2 \\ -m_1&q&m_2
	\end{pmatrix}
\]
and that the tensor product of spherical tensors takes the form (where $[J]=2J+1$)
\begin{equation}
    [A\otimes B]^{Kq}=\sum_{m_{a,b}}\sqrt{[K]}(-)^{K-q}\begin{pmatrix}
		1 & 1 & K \\m_a&m_b&-q
	\end{pmatrix}A^{m_a}B^{m_b}
\end{equation} 
A tedious but straightforward calculation then shows that
\begin{widetext}
	\begin{align}
		\braket{Jm_\alpha|\alpha^{K,q}|Jm_\beta} & =\nonumber\sqrt{[K]}\sum_\gamma(-)^{m_a+m_b}\begin{pmatrix}1&1&K\\-m_a&-m_b &q\end{pmatrix}\frac{\braket{Jm_\alpha|d_{m_a}|J_\gamma m_\gamma}\braket{J_\gamma m_\gamma|d_{m_b}|J m_\beta}}{\omega_\gamma-\omega}                                                                 \\
		                                         & =\nonumber\sqrt{[K]}\sum_\gamma(-)^{m_a+m_b-m_\alpha+m_\gamma}\frac{|\braket{J||\vb d||J_\gamma}|^2}{\omega_\gamma-\omega}\begin{pmatrix}1&1&K\\-m_a&-m_b&q\end{pmatrix}\begin{pmatrix}J&1&J_\gamma\\-m_\alpha&m_a&m_\gamma\end{pmatrix}\begin{pmatrix}
			                                                                                                                                                                                                                                                                          J_\gamma & 1 & J \\ -m_\gamma&m_b&m_\beta
		                                                                                                                                                                                                                                                                          \end{pmatrix} \\
		                                         & =\sqrt{[K]}\sum_\gamma(-)^{J_\gamma-m_\alpha}\frac{|\braket{J||\vb d||J_\gamma}|^2}{\omega_\gamma-\omega}\begin{Bmatrix}J&J&K\\1&1&J_\gamma\end{Bmatrix}\begin{pmatrix}J&J&K\\m_\beta&-m_\alpha&q\end{pmatrix}
		\end{align}
        Hence the reduced matrix elements of the polarizability take the form
        \begin{equation}
		\braket{J||\alpha^K||J}                   =\sum_{\gamma}\sqrt{[K]}(-)^{J_\gamma-J}\frac{|\braket{J||\vb d||J_\gamma}|^2}{\omega_\gamma-\omega}\begin{Bmatrix}J&J&K\\1&1&J_\gamma\end{Bmatrix}                                                                                                                   \end{equation}
        Of particular interest to the alkali metals is the $J=1/2$ case, which represents the ground state
        \begin{align}
		\braket{1/2||\alpha^K||1/2}              & =\sum_{\gamma}\frac{|\braket{1/2||\vb d||J_\gamma}|^2}{\omega_\gamma-\omega}\times \begin{cases}
			                                                                                                                              \frac{1}{\sqrt{6}}                          & K=0 \\
			                                                                                                                              -\frac{11-4J_\gamma(J_\gamma+1)}{8\sqrt{3}} & K=1
		                                                                                                                              \end{cases}                                                                                                                \\
		                                         & =\sum_{\gamma}\frac{|\braket{1/2||\vb d||J_\gamma}|^2}{\omega_\gamma-\omega}\times \begin{cases}
			                                                                                                                              \frac{\braket{1/2||1||1/2}}{2\sqrt{3}}                                & K=0 \\
			                                                                                                                              \frac{11-4J_\gamma(J_\gamma+1)}{12\sqrt{2}}\i\braket{1/2||\vb J||1/2} & K=1
		                                                                                                                              \end{cases}
	\end{align}
	To rewrite this in the usual form, we note that with our convention
	\begin{align}
		[\bar{\vb E}(\omega_1)\otimes \vb E(\omega_2)]^{Kq} & =\begin{cases}
			                                                       \frac{\bar{\vb E}(\omega_1)\cdot\vb E(\omega_2)}{\sqrt{3}}      & K=0 \\
			                                                       \frac{[\bar{\vb E}(\omega_1)\times\vb E(\omega_2)]_q}{\sqrt{2}} & K=1
		                                                       \end{cases}
	\end{align}
    For alkali metals, the dominant sources of polarizability are the $D_{1,2}$ lines connecting the ground state $S_{1/2}$ to the $P_{1/2,3/2}$ levels at energy $\omega_{1/2,3/2}$, where $\braket{S_{1/2}|\vb d|P_{3/2}}=\sqrt{2}\braket{S_{1/2}|\vb d|P_{1/2}}=\sqrt{2}d$. Hence the second order Hamiltonian takes the form
	\begin{align}
		\delta H^{(2)} & =-\sum_{\gamma,\omega_{1,2}\,\mr{res}} \frac{|\braket{S_{1/2}||\vb d||J_\gamma}|^2}{\omega_\gamma-\omega}\left[\frac{\bar{\vb E}(\omega_1)\cdot\vb E(\omega_2)}{6}+\frac{\i(11-4J_\gamma(J_\gamma+1))\vb J\cdot [\bar{\vb E}(\omega_1)\times \vb E(\omega_2)]}{24}\right]                                   \\
		               & =\sum_{\omega_{1,2}\,\mr{res}} \frac{d^2\bar{\vb E}(\omega_1)\cdot\vb E(\omega_2)}{6}\left[\frac{1}{\omega-\omega_{1/2}}+\frac{2}{\omega-\omega_{3/2}}\right]+\frac{\i d^2\vb J\cdot \bar{\vb E}(\omega_1)\times\vb E(\omega_2)}{3}\left[\frac{1}{\omega-\omega_{1/2}}-\frac{1}{\omega-\omega_{3/2}}\right]\,.
	\end{align}
\end{widetext}
The most important feature is that the vector polarizability from the two lines destructively add when far detuned. In this regime, when  $\omega\ll  \omega_{1/2},\omega_{3/2}$ or $\omega\gg \omega_{1/2},\omega_{3/2}$, we find \begin{equation}
    \alpha_V\to\frac{ d^2 (\omega_{3/2}-\omega_{1/2})}{3(\omega-\omega_{1/2})^2}.
\end{equation}

As discussed, the imaginary part of the polarizability can be obtained by $\omega_{1/2,3/2}\to\omega_{1/2,3/2}-\i\Gamma/2$. When far detuned the imaginary part of the scalar potential scales in the same way as $\alpha_V$, namely
\begin{equation}
  \mr{Im} \alpha_S\to \frac{d^2\Gamma}{4(\omega-\omega_{1/2})^2}.
  \end{equation}
   \begin{table}[h!]
\renewcommand{\arraystretch}{1.4}
 \begin{tabular}{r|ddddd}
\toprule
    Element & {\mr{Li}} & \mr{Na}& \mr{K}&\mr{Rb}&\mr{Cs} \\ 
    \midrule
    ${\Gamma}/{2\pi\,\si{\mega\hertz}}$&5.9&9.8&6.0&5.9&4.9\\
    $({\omega_{3/2}-\omega_{1/2}})/{2\pi\,\si{\tera\hertz}}$&0.01&0.52&1.7&7.1&16.6\\
      ${\Im \alpha_S}/{\alpha_V}\times 10^{4} $&4&0.1&0.03&0.006&0.003\\     
    \bottomrule
\end{tabular}
\renewcommand{\arraystretch}{1}
    \caption{Key spectral features of $\mathrm{D}_{1,2}$ lines in alkali metals. Splitting data from \cite{Lifreq,Nafreq,Kfreq,Rbfreq,CsD1,CsD2} and lifetime data from \cite{volz1996precision,Cslifetime}. At the stated precision the isotope shift is negligible, and we have quoted the average $\Gamma$ for the two lines. The quoted ratio $\Im\alpha_S/\alpha_V$ is the asymptotically approached ratio in the far detuned regime.}
    \label{tab:Dline}
\end{table}

  The ratio of these two quantities is a key Fig. of merit for the lifetime of alkali metals in OFLs
\begin{equation}
	\frac{\mr{Im}\alpha_S}{\alpha_V}\to \frac{3\Gamma}{4(\omega_{3/2}-\omega_{1/2})}.
\end{equation}
In particular, given some fixed non-scalar potential, no detuning will give lower heating than that suggested by taking this ratio to its asymptotic value. In practice it is a good estimate so long as $\omega$ is a multiple of $\omega_{3/2}-\omega_{1/2}$ from resonance. This figure of merit explains why heavy alkali-metals are preferable, as the fine structure splitting $\omega_{3/2}-\omega_{1/2}$ is a relativistic effect that increases with atomic number. We quote the relevant values in Table \ref{tab:Dline}.

\end{document}